% Quenched vs dynamic Coulomb gauge wave functions
%\def\preprint{Y}        %Y for preprint version
\def\preprint{N}        %N for  non-preprint version
%\input jnl
%\input figuredefs
%\input reforder
%\input tables.tex
%%                      	JNL.TEX

%%
%%                This is JNL.TEX Version 0.3 as of 6/12/85.
%%
%%	This is a set of TeX 82 macros designed to produce scientific
%%	papers with a minimum of fuss and using as much of plain.tex as
%%	possible.  The user need only know what is in the TeXbook, and
%%	the macros under ``user definitions'' below.  Also, the user
%%	definitions are intended to be as simple as possible, so that
%%	the user may change them as desired.

%%
%%  Font definitions suitable for the IMAGEN (Written by Tony Kennedy)
%%

%  Define a whole menagerie of pseudo-12pt fonts

\font\twelverm=cmr10 scaled 1200    \font\twelvei=cmmi10 scaled 1200
\font\twelvesy=cmsy10 scaled 1200   \font\twelveex=cmex10 scaled 1200
\font\twelvebf=cmbx10 scaled 1200   \font\twelvesl=cmsl10 scaled 1200
\font\twelvett=cmtt10 scaled 1200   \font\twelveit=cmti10 scaled 1200

\skewchar\twelvei='177   \skewchar\twelvesy='60

%  Define \...point macros to change fonts and spacings consistently

\def\twelvepoint{\normalbaselineskip=12.4pt
  \abovedisplayskip 12.4pt plus 3pt minus 9pt
  \belowdisplayskip 12.4pt plus 3pt minus 9pt
  \abovedisplayshortskip 0pt plus 3pt
  \belowdisplayshortskip 7.2pt plus 3pt minus 4pt
  \smallskipamount=3.6pt plus1.2pt minus1.2pt
  \medskipamount=7.2pt plus2.4pt minus2.4pt
  \bigskipamount=14.4pt plus4.8pt minus4.8pt
  \def\rm{\fam0\twelverm}          \def\it{\fam\itfam\twelveit}%
  \def\sl{\fam\slfam\twelvesl}     \def\bf{\fam\bffam\twelvebf}%
  \def\mit{\fam 1}                 \def\cal{\fam 2}%
  \def\tt{\twelvett}
  \textfont0=\twelverm   \scriptfont0=\tenrm   \scriptscriptfont0=\sevenrm
  \textfont1=\twelvei    \scriptfont1=\teni    \scriptscriptfont1=\seveni
  \textfont2=\twelvesy   \scriptfont2=\tensy   \scriptscriptfont2=\sevensy
  \textfont3=\twelveex   \scriptfont3=\twelveex  \scriptscriptfont3=\twelveex
  \textfont\itfam=\twelveit
  \textfont\slfam=\twelvesl
  \textfont\bffam=\twelvebf \scriptfont\bffam=\tenbf
  \scriptscriptfont\bffam=\sevenbf
  \normalbaselines\rm}

%	tenpoint

%%
%%	Various internal macros
%%

\def\beginlinemode{\endmode
  \begingroup\parskip=0pt \obeylines\def\\{\par}\def\endmode{\par\endgroup}}
\def\beginparmode{\endmode
  \begingroup \def\endmode{\par\endgroup}}
\let\endmode=\par
{\obeylines\gdef\
{}}
\def\singlespace{\baselineskip=\normalbaselineskip}
\def\oneandathirdspace{\baselineskip=\normalbaselineskip
  \multiply\baselineskip by 4 \divide\baselineskip by 3}
\def\oneandahalfspace{\baselineskip=\normalbaselineskip
  \multiply\baselineskip by 3 \divide\baselineskip by 2}
\def\doublespace{\baselineskip=\normalbaselineskip \multiply\baselineskip by 2}

\newcount\firstpageno
\firstpageno=2
\footline={\ifnum\pageno<\firstpageno{\hfil}\else{\hfil\twelverm\folio\hfil}\fi}
\let\rawfootnote=\footnote		% We must set the footnote style
\def\footnote#1#2{{\rm\singlespace\parindent=0pt\rawfootnote{#1}{#2}}}
\def\raggedcenter{\leftskip=4em plus 12em \rightskip=\leftskip
  \parindent=0pt \parfillskip=0pt \spaceskip=.3333em \xspaceskip=.5em
  \pretolerance=9999 \tolerance=9999
  \hyphenpenalty=9999 \exhyphenpenalty=9999 }
\def\dateline{\rightline{\ifcase\month\or
  January\or February\or March\or April\or May\or June\or
  July\or August\or September\or October\or November\or December\fi
  \space\number\year}}
\def\received{\vskip 3pt plus 0.2fill
 \centerline{\sl (Received\space\ifcase\month\or
  January\or February\or March\or April\or May\or June\or
  July\or August\or September\or October\or November\or December\fi
  \qquad, \number\year)}}

%%
%%	Page layout, margins, font and spacing (feel free to change)
%%

\hsize=6.5truein
\hoffset=0truein
\vsize=8.9truein
\voffset=0truein
\parskip=\medskipamount
\twelvepoint		% selects twelvepoint fonts (cf. \tenpoint)
\doublespace		% selects double spacing for main part of paper (cf.
			%	\singlespace, \oneandahalfspace)
\overfullrule=0pt	% delete the nasty little black boxes for overfull box

%%
%%	The user definitions for major parts of a paper (feel free to change)
%%

\def\preprintno#1{
 \rightline{\rm #1}}	% Preprint number at upper right of title page

\def\title			%  Title on title page
  {\null\vskip 3pt plus 0.2fill
   \beginlinemode \doublespace \raggedcenter \bf}

\def\author			%  Author(s) name(s)  on title page
  {\vskip 3pt plus 0.2fill \beginlinemode
   \singlespace \raggedcenter}

\def\affil			% Affiliations (can intermix with \author)
  {\vskip 3pt plus 0.1fill \beginlinemode
   \oneandahalfspace \raggedcenter \sl}

\def\abstract			% Begin abstract
  {\vskip 3pt plus 0.3fill \beginparmode
   \doublespace \narrower ABSTRACT: }

\def\endtitlepage		% End title page, begin body of paper
  {\endpage			% 	This subsumes \body
   \body}

\def\body			% Begin text body;  can be used to end
  {\beginparmode}		% \title, \author, \affil, \abstract,
				% \reference, or \figurecaption modes

\def\head#1{			% Head;  NOTE enclose the text in {}
  \filbreak\vskip 0.5truein	%  e.g., \head{I. Introduction}
  {\immediate\write16{#1}
   \raggedcenter \uppercase{#1}\par}
   \nobreak\vskip 0.25truein\nobreak}

\def\subhead#1{			% Subhead;  NOTE enclose the text in {}
  \vskip 0.25truein		% e.g., \subhead{A. History of the Problem}
  {\raggedcenter #1 \par}
   \nobreak\vskip 0.25truein\nobreak}

\def\refto#1{$^{#1}$}		% For references in text as superscript

\def\references			% Begin references -- basic format is Phys Rev
  {\head{References}		% I.e., volume, page, year (space after commas).
   \beginparmode
   \frenchspacing \parindent=0pt \leftskip=1truecm
   \parskip=8pt plus 3pt \everypar{\hangindent=\parindent}}

\gdef\refis#1{\indent\hbox to 0pt{\hss#1.~}}	% Ref list numbers.

\gdef\journal#1, #2, #3, 1#4#5#6{		% Journal reference.  Comma sets
    {\sl #1~}{\bf #2}, #3 (1#4#5#6)}		% off: name, vol, page, year

\gdef\journ2 #1, #2, #3, 1#4#5#6{		% Journal reference.  Comma sets
    {\sl #1~}{\bf #2}: #3 (1#4#5#6)}		% off: name, vol, page, year
                                     		% Colon inserted after volume #

\def\refstylenp{		% Nucl Phys(or Phys Lett) ref style: V, Y, P
  \gdef\refto##1{ [##1]}				% Reference in text []
  \gdef\refis##1{\indent\hbox to 0pt{\hss##1)~}}	% Ref list numbers)
  \gdef\journal##1, ##2, ##3, ##4 {			% Journal reference
     {\sl ##1~}{\bf ##2~}(##3) ##4 }}

\def\refstyleprnp{		% Input like pr, output like np!!
  \gdef\refto##1{ [##1]}				% Reference in text []
  \gdef\refis##1{\indent\hbox to 0pt{\hss##1)~}}	% Ref list numbers)
  \gdef\journal##1, ##2, ##3, 1##4##5##6{		% Journal reference
    {\sl ##1~}{\bf ##2~}(1##4##5##6) ##3}}

\def\prd{\journal Phys. Rev. D, }

\def\prl{\journal Phys. Rev. Lett., }

\def\np{\journal Nucl. Phys., }

\def\pl{\journal Phys. Lett., }

\def\endreferences{\body}

\def\figurecaptions		% Begin figure captions
  {\endpage
   \beginparmode
   \head{Figure Captions}
}

\def\endfigurecaptions{\body}

\def\endpage			%  Eject a page
  {\vfill\eject}

\def\endpaper			%  Ways to say goodbye
  {\endmode\vfill\supereject}

\def\endit
  {\endpaper\end}

%%
%%	Various little user definitions
%%

\def\ref#1{Ref. #1}			% 	for inline references
\def\Ref#1{Ref. #1}			% 	ditto

		% For citation of equation numbers
	%	ditto
			%	ditto
			%	ditto
			%	ditto
			%	ditto
\def\frac#1#2{{\textstyle #1 \over \textstyle #2}}

\def\sla{\raise.15ex\hbox{$/$}\kern-.57em}
\def\leaderfill{\leaders\hbox to 1em{\hss.\hss}\hfill}
\def\twiddle{\lower.9ex\rlap{$\kern-.1em\scriptstyle\sim$}}
\def\bigtwiddle{\lower1.ex\rlap{$\sim$}}
\def\gtwid{\mathrel{\raise.3ex\hbox{$>$\kern-.75em\lower1ex\hbox{$\sim$}}}}
\def\ltwid{\mathrel{\raise.3ex\hbox{$<$\kern-.75em\lower1ex\hbox{$\sim$}}}}
\def\square{\kern1pt\vbox{\hrule height 1.2pt\hbox{\vrule width 1.2pt\hskip 3pt
   \vbox{\vskip 6pt}\hskip 3pt\vrule width 0.6pt}\hrule height 0.6pt}\kern1pt}

\def\begintable{\offinterlineskip\hrule}
\def\endtable{\hrule}

\catcode`@=11
\newcount\r@fcount \r@fcount=0
\newcount\r@fcurr
\immediate\newwrite\reffile
\newif\ifr@ffile\r@ffilefalse
\def\w@rnwrite#1{\ifr@ffile\immediate\write\reffile{#1}\fi\message{#1}}

\def\writer@f#1>>{}
\def\referencefile{%			  Stuff to write .REF file
  \r@ffiletrue\immediate\openout\reffile=\jobname.ref%
  \def\writer@f##1>>{\ifr@ffile\immediate\write\reffile%
    {\noexpand\refis{##1} = \csname r@fnum##1\endcsname = %
     \expandafter\expandafter\expandafter\strip@t\expandafter%
     \meaning\csname r@ftext\csname r@fnum##1\endcsname\endcsname}\fi}%
  \def\strip@t##1>>{}}

\def\citeall#1{\xdef#1##1{#1{\noexpand\cite{##1}}}}
\def\cite#1{\each@rg\citer@nge{#1}}	% Variable No. of args, separated by ","

\def\each@rg#1#2{{\let\thecsname=#1\expandafter\first@rg#2,\end,}}
\def\first@rg#1,{\thecsname{#1}\apply@rg}	% each@ag is a general purpose
\def\apply@rg#1,{\ifx\end#1\let\next=\relax%	  variable no. of arg. macro.
\else,\thecsname{#1}\let\next=\apply@rg\fi\next}% args separated by commas

\def\citer@nge#1{\citedor@nge#1-\end-}	% Check for M-N range (M and N numbers)
\def\citer@ngeat#1\end-{#1}
\def\citedor@nge#1-#2-{\ifx\end#2\r@featspace#1 % Single argument
  \else\citel@@p{#1}{#2}\citer@ngeat\fi}	% M-N range of arguments
\def\citel@@p#1#2{\ifnum#1>#2{\errmessage{Reference range #1-#2\space is bad.}%
    \errhelp{If you cite a series of references by the notation M-N, then M and
    N must be integers, and N must be greater than or equal to M.}}\else%
 {\count0=#1\count1=#2\advance\count1
by1\relax\expandafter\r@fcite\the\count0,%
  \loop\advance\count0 by1\relax%	  Loop from M to N
    \ifnum\count0<\count1,\expandafter\r@fcite\the\count0,%
  \repeat}\fi}

\def\r@featspace#1#2 {\r@fcite#1#2,}	% Eat spaces at beginning or end of arg
\def\r@fcite#1,{\ifuncit@d{#1}%		  Cite individual reference
    \newr@f{#1}%
    \expandafter\gdef\csname r@ftext\number\r@fcount\endcsname%
                     {\message{Reference #1 to be supplied.}%
                      \writer@f#1>>#1 to be supplied.\par}%
 \fi%
 \csname r@fnum#1\endcsname}
\def\ifuncit@d#1{\expandafter\ifx\csname r@fnum#1\endcsname\relax}%
\def\newr@f#1{\global\advance\r@fcount by1%
    \expandafter\xdef\csname r@fnum#1\endcsname{\number\r@fcount}}

\let\r@fis=\refis			% Save old \refis, redefine
\def\refis#1#2#3\par{\ifuncit@d{#1}%      Use two params #2 #3 to strip blank
   \newr@f{#1}%
   \w@rnwrite{Reference #1=\number\r@fcount\space is not cited up to now.}\fi%
  \expandafter\gdef\csname r@ftext\csname r@fnum#1\endcsname\endcsname%
  {\writer@f#1>>#2#3\par}}

\def\ignoreuncited{%   redefine \refis if ignoring uncited references
   \def\refis##1##2##3\par{\ifuncit@d{##1}%
     \else\expandafter\gdef\csname r@ftext\csname
r@fnum##1\endcsname\endcsname%
     {\writer@f##1>>##2##3\par}\fi}}

\def\r@ferr{\endreferences\errmessage{I was expecting to see
\noexpand\endreferences before now;  I have inserted it here.}}
\let\r@ferences=\references
\def\references{\r@ferences\def\endmode{\r@ferr\par\endgroup}}

\let\endr@ferences=\endreferences
\def\endreferences{\r@fcurr=0%		  Save old \endreferences, redefine
  {\loop\ifnum\r@fcurr<\r@fcount%	  Loop over refnum and produce text
    \advance\r@fcurr by 1\relax\expandafter\r@fis\expandafter{\number\r@fcurr}%
    \csname r@ftext\number\r@fcurr\endcsname%
  \repeat}\gdef\r@ferr{}\endr@ferences}

% Save old \endpaper, redefine it to write parting message.

\let\r@fend=\endpaper\gdef\endpaper{\ifr@ffile
\immediate\write16{Cross References written on []\jobname.REF.}\fi\r@fend}

\catcode`@=12

\citeall\refto		% These macros will generate citations
\citeall\ref		%
\citeall\Ref		%

% +--------------------------------------------------------------------+
% |                                                                    |
% |                           TABLES.TEX                               |
% |                                                                    |
% |                     Ray F. Cowan  15-Feb-85                        |
% |                                                                    |
% |                       Princeton University                         |
% |                                                                    |
% |                     Last Revision: 21-Nov-85                       |
% |                                                                    |
% |   Macros I find handy for making tables.  See TABLEDOC TEX for     |
% |   a longer description.  The token-counting macros are straight    |
% |   from the TeXbook's "Dirty Tricks" appendix.                      |
% |                                                                    |
% +--------------------------------------------------------------------+
%
\newbox\hdbox%
\newcount\hdrows%
\newcount\multispancount%
\newcount\ncase%
\newcount\ncols% This is the number of primary text columns in the table.
\newcount\nrows%
\newcount\nspan%
\newcount\ntemp%
\newdimen\hdsize%
\newdimen\newhdsize%
\newdimen\parasize%
\newdimen\spreadwidth%
\newdimen\thicksize%
\newdimen\thinsize%
\newdimen\tablewidth%
\newif\ifcentertables%
\newif\ifendsize%
\newif\iffirstrow%
\newif\iftableinfo%
\newtoks\dbt%
\newtoks\hdtks%
\newtoks\savetks%
\newtoks\tableLETtokens%
\newtoks\tabletokens%
\newtoks\widthspec%
%
%  Book-keeping stuff--see how often these macros are called.
%
%  Turn on table diagnostics.
%
\tableinfotrue%
\catcode`\@=11%  Allows use of "@" in macro names, like PLAIN.TEX does.
%  Debugging aid.  Writes #1 on the
%                                    user's terminal and in the log file.
%
%  Define the \tstrut height, depth in terms of the x_height parameter.
%
\def\tstrut{\vrule height3.1ex depth1.2ex width0pt}%
\def\and{\char`\&}%  Allows us to get an `&' in the text.  This is the
%                    same as using the PLAIN TeX macro \&.
\def\tablerule{\noalign{\hrule height\thinsize depth0pt}}%
\thicksize=.6pt%  Default thickness for fat rules.  The user should feel
%                  free to change this to his preference.
\thinsize=0.6pt%   Default thickness for thin rules.
\def\thickrule{\noalign{\hrule height\thicksize depth0pt}}%
\def\ctr#1{\hfil\ #1\hfil}%
%
%
%
%  Here are things for controlling the width of the finished table.
%
\tablewidth=-\maxdimen%
\spreadwidth=-\maxdimen%
\def\tabskipglue{0pt plus 1fil minus 1fil}%
%
%  Stuff for centering or not.
%
\centertablestrue%
%
%
%
%  \vctr vertically centers its argument in the row.
%
\parasize=4in%
\gdef\ARGS{########}%  Produces the correct number of #'s in the preamble
%                      by the time eveything is expanded and \halign sees
%                      it.
\gdef\headerARGS{####}%  Same as \ARGS, but used in \header macros.
\def\@mpersand{&}%  Allows us to get alignment tab characters later
%                   when we have made the character "&" an active macro.
{\catcode`\|=13%  Make |'s locally active.
\gdef\letbarzero{\let|0}%  Globally define a macro that allows us to
%                          keep active |'s from being expanded in edef's.
\gdef\letbartab{\def|{&&}}%
\gdef\letvbbar{\let\vb|}%
%  This \def will cause active |'s read by
%                            \ruledtable to be converted into double
%                            alignment tabs.
}%  End of locally active |'s.
{\catcode`\&=4%  Make these alignment tabs.
\def\ampskip{&\omit\hfil&}%  This local macro skips a vertical rule.
\catcode`\&=13%  Now make &'s into active macros.
\let&0%  This allows us to expand \ampskip in the next \xdef without
%        attempting to expand the & and getting an "undefined control
%        sequence" error.
\xdef\letampskip{\def&{\ampskip}}%
\gdef\letnovbamp{\let\novb&\let\tab&}
%  This will cause active &'s read by
%                                   \ruledtable to be converted into
%                                   double tabs and an \omit'ted \vrule.
}%  End of locally active &'s.
\def\begintable{%  Here we make |'s and &'s active characters so we can
%                  interpret them as macros.  Note that this action is
%                  true only until we encounter the matching \endgroup
%                  token later at the end of the \ruledtable macro.
   \begingroup%
   \catcode`\|=13\letbartab\letvbbar%
   \catcode`\&=13\letampskip\letnovbamp%
   \def\multispan##1{%  We must redefine \multispan to count the number
%                       of primary columns, not physical columns.
      \omit \mscount##1%
      \multiply\mscount\tw@\advance\mscount\m@ne%
      \loop\ifnum\mscount>\@ne \sp@n\repeat%
   }%  End of \multispan macro.
   \def\|{%
      &\omit\widevline&%
   }%
   \ruledtable%  Now we call \ruledtable to do the real work.
}%  End of \begintable macro.
\long\def\ruledtable#1\endtable{%
%
%  This macro reads in the user's data entries
%  and converts them into a ruled table.
%
%  Important note:  Many macros and parameters are re-defined here, and
%  these must be kept local to the table macros to avoid conflict with
%  their use outside of tables.  This is done by the \begingroup token
%  macro \begintable and the \endgroup token at the end of
%  this macro.
%
   \offinterlineskip%  Needed to make rules touch each other.
   \tabskip 0pt%  Needed for same reason as \offinterlineskip.
   \def\widevline{\vrule width\thicksize}%  Make outer \vrule's wider.
   \def\endrow{\@mpersand\omit\hfil\crnorm\@mpersand}%
   \def\crthick{\@mpersand\crnorm\thickrule\@mpersand}%
   \def\crnorule{\@mpersand\crnorm\@mpersand}%
   \let\nr=\crnorule%  A shorter abbreviation.
   \def\endtable{\@mpersand\crnorm\thickrule}%
   \let\crnorm=\cr%  Allows us to use \cr for our own purposes.
%
%  Cause user-typed \cr's to follow a row with a \tablerule.
%
   \edef\cr{\@mpersand\crnorm\tablerule\@mpersand}%
   \the\tableLETtokens%  Get the user's extra \let's, if any.
%
%  Put the data entries into a token register so we can scan through them
%  and see what the user is asking us to do.
%
   \tabletokens={&#1}%  We add an extra alignment tab to the beginning
%                       of the first row to allow for the first \vrule.
%
%  Now count how many rows are in the table and return the result in
%  count register \nrows; do the same for columns, and return that
%  in register \ncols.
%
   \countROWS\tabletokens\into\nrows%
   \countCOLS\tabletokens\into\ncols%
%
%  Now do a little arithmetic to convert the number of primary columns
%  into the number of physical columns that the alignment preamble must
%  prepare for;  similarly for rows.
%
   \advance\ncols by -1%
   \divide\ncols by 2%
   \advance\nrows by 1%
%
%  Tell the user how many rows and columns we found in his data, if he
%  wants to know.
%
   \iftableinfo %
      \immediate\write16{[Nrows=\the\nrows, Ncols=\the\ncols]}%
   \fi%
%
%  Now we actually go ahead and produce the table.
%
   \ifcentertables
      \ifhmode \par\fi%  Make sure we are in vertical mode.
      \line{%  The final table comes out as an \hbox of width the \hsize.
      \hss%  The final table will be centered left-to-right.
   \else %
      \hbox{%
   \fi
      \vbox{%
         \makePREAMBLE{\the\ncols}%  Generate the preamble.
         \edef\next{\preamble}%  This line and the next line force the
         \let\preamble=\next%    expansion of all \ARGS tokens into the
%                                appropriate number of #'s.
         \makeTABLE{\preamble}{\tabletokens}%  Go do the \halign here.
      }%  End of \vbox.
      \ifcentertables \hss}\else }\fi%  Finish the centering effect.
%                                       It is important that no spaces
%                                       follow the two `}' here.
%  }%  End of \line.
   \endgroup%  Return all local macros and parameters to their outside
%              values.
   \tablewidth=-\maxdimen%  Reset \tablewidth to normal.
   \spreadwidth=-\maxdimen% Same for \spreadwidth.
}%  End of macro \ruledtable.
\def\makeTABLE#1#2{%  Does an \halign for the \ruledtable macro.
   {%  Start of local parameter values.
   \let\ifmath0%     These macros would cause trouble if they were to be
   \let\header0%     expanded in the following \xdef; we \let them be
   \let\multispan0%  equal to a digit, because digits can't be expanded.
%
%  Set up the width specification here.
%
   \ncase=0%
   \ifdim\tablewidth>-\maxdimen \ncase=1\fi%
   \ifdim\spreadwidth>-\maxdimen \ncase=2\fi%
   \relax%  This \relax is absolutely necessary, without it the following
%           \ifcase will always take \ncase=0.
%
   \ifcase\ncase %
      \widthspec={}%
   \or %
      \widthspec=\expandafter{\expandafter t\expandafter o%
                 \the\tablewidth}%
   \else %
      \widthspec=\expandafter{\expandafter s\expandafter p\expandafter r%
                 \expandafter e\expandafter a\expandafter d%
                 \the\spreadwidth}%
   \fi %
%\out{Widthspec=[\the\widthspec]}%
%\out{Preamble=[\preamble]}%
   \xdef\next{%  We must force the preamble to be expanded BEFORE the
      \halign\the\widthspec{%
%        \halign is done;  this \edef\next{...}\next construction
%                does the trick.
      #1%  This is the preamble text.
      \noalign{\hrule height\thicksize depth0pt}%  Makes the top \hrule.
      \the#2\endtable%  This is the main body.
%
%     \noalign{\hrule height0.7pt depth0pt}%  Makes the last \hrule.
      }%  End of \halign.
   }%  End of \next.
   }%  End of local values.
   \next%  This \next must be outside of the local values, because now
%          we want those troublesome macros in the \let's above to have
%          their normal actions.
}%  End of macro \makeTABLE.
\def\makePREAMBLE#1{%  This macro generates the necessary preamble for a
%                      ruled table with #1 primary columns.
%                      (Primary columns means the number of columns NOT
%                       counting those used for vertical rules.)
   \ncols=#1%  Get the number of columns desired.
   \begingroup%  Start local parameter definitions.
   \let\ARGS=0%  This is the key to the whole thing; it prevents \ARGS
%                from being expanded in the following \edef's.
   \edef\xtp{\widevline\ARGS\tabskip\tabskipglue%
   &\ctr{\ARGS}\tstrut}%  A 1-column preamble.  Gets the sizing right.
   \advance\ncols by -1%  One column has been generated; decrement the
%                         counter.
   \loop%  Append as many further columns as needed to the preamble.
      \ifnum\ncols>0 %
      \advance\ncols by -1%
      \edef\xtp{\xtp&\vrule width\thinsize\ARGS&\ctr{\ARGS}}%
   \repeat
   \xdef\preamble{\xtp&\widevline\ARGS\tabskip0pt%
   \crnorm}%  Adds the last \vrule.
   \endgroup%  End of local parameters.
}%  End of macro \makePREAMBLE.
\def\countROWS#1\into#2{%  This counts the number of rows in #1 by
%                          looking for control sequences that end a row,
%                          e.g., \cr, \crthick, etc., and puts the result
%                          into count register #2.
   \let\countREGISTER=#2%
   \countREGISTER=0%
%  \out{In countROWS:  tokens are [\the#1]}%
   \expandafter\ROWcount\the#1\endcount%
}%
\def\ROWcount{%
   \afterassignment\subROWcount\let\next= %
}%
\def\subROWcount{%
%  \out{In subROWcount:  next is [\meaning\next]}%  Debugging aid.
   \ifx\next\endcount %
      \let\next=\relax%
   \else%
      \ncase=0%
      \ifx\next\cr %
         \global\advance\countREGISTER by 1%
         \ncase=0%
      \fi%
      \ifx\next\endrow %
         \global\advance\countREGISTER by 1%
         \ncase=0%
      \fi%
      \ifx\next\crthick %
         \global\advance\countREGISTER by 1%
         \ncase=0%
      \fi%
      \ifx\next\crnorule %
         \global\advance\countREGISTER by 1%
         \ncase=0%
      \fi%
      \ifx\next\header %
%     \out{In subROWcount:  next=header, ncase set=1}%
         \ncase=1%
      \fi%
%     \out{In subROWcount:  ncase is [\the\ncase]}%
      \relax%
      \ifcase\ncase %
         \let\next\ROWcount%
%        \out{subROWcount---> ncase=\the\ncase}%
      \or %
         \let\next\argROWskip%
%        \out{subROWcount---> ncase=\the\ncase}%
      \else %
      \fi%
   \fi%
%  \out{subROWcount---> NEXT=\meaning\next}%
   \next%
}%  End of macro \subROWcount.
\def\counthdROWS#1\into#2{%
\dvr{10}%
   \let\countREGISTER=#2%
   \countREGISTER=0%
\dvr{11}%
%  \out{In counthdROWS:  tokens are [\the#1]}%
\dvr{13}%
   \expandafter\hdROWcount\the#1\endcount%
\dvr{12}%
}%
\def\hdROWcount{%
   \afterassignment\subhdROWcount\let\next= %
}%
\def\subhdROWcount{%
%\out{In subhdROWcount:  next is [\meaning\next]}%
   \ifx\next\endcount %
      \let\next=\relax%
   \else%
      \ncase=0%
      \ifx\next\cr %
         \global\advance\countREGISTER by 1%
         \ncase=0%
      \fi%
      \ifx\next\endrow %
         \global\advance\countREGISTER by 1%
         \ncase=0%
      \fi%
      \ifx\next\crthick %
         \global\advance\countREGISTER by 1%
         \ncase=0%
      \fi%
      \ifx\next\crnorule %
         \global\advance\countREGISTER by 1%
         \ncase=0%
      \fi%
      \ifx\next\header %
%\out{In subhdROWcount:  next=header, ncase set=1}%
         \ncase=1%
      \fi%
%\out{In subhdROWcount:  ncase is [\the\ncase]}%
\relax%
      \ifcase\ncase %
         \let\next\hdROWcount%
%\out{subhdROWcount---> ncase=\the\ncase}%
      \or%
         \let\next\arghdROWskip%
%\out{subhdROWcount---> ncase=\the\ncase}%
      \else %
      \fi%
   \fi%
%\out{subhdROWcount---> NEXT=\meaning\next}%
   \next%
}%
{\catcode`\|=13\letbartab
\gdef\countCOLS#1\into#2{%
%  \out{In countCOLS:  tokens are [\the#1]}
   \let\countREGISTER=#2%
   \global\countREGISTER=0%
   \global\multispancount=0%
   \global\firstrowtrue
   \expandafter\COLcount\the#1\endcount%
   \global\advance\countREGISTER by 3%
   \global\advance\countREGISTER by -\multispancount
%  \out{countCOLS-->[\the\countREGISTER]}
}%
\gdef\COLcount{%
   \afterassignment\subCOLcount\let\next= %
}%
{\catcode`\&=13%
\gdef\subCOLcount{%
%\out{In subCOLcount: next is [\meaning\next]}
   \ifx\next\endcount %
      \let\next=\relax%
   \else%
      \ncase=0%
      \iffirstrow
         \ifx\next& %
            \global\advance\countREGISTER by 2%
            \ncase=0%
         \fi%
         \ifx\next\span %
            \global\advance\countREGISTER by 1%
            \ncase=0%
         \fi%
         \ifx\next| %
            \global\advance\countREGISTER by 2%
            \ncase=0%
         \fi
         \ifx\next\|
            \global\advance\countREGISTER by 2%
            \ncase=0%
         \fi
         \ifx\next\multispan
            \ncase=1%
            \global\advance\multispancount by 1%
         \fi
         \ifx\next\header
            \ncase=2%
         \fi
         \ifx\next\cr       \global\firstrowfalse \fi
         \ifx\next\endrow   \global\firstrowfalse \fi
         \ifx\next\crthick  \global\firstrowfalse \fi
         \ifx\next\crnorule \global\firstrowfalse \fi
      \fi%  End of \iffirstrow.
\relax%\out{subCOL-->  ncase=[\the\ncase]}
% \out{subCOL-->  next=\meaning\next}
      \ifcase\ncase %
         \let\next\COLcount%
      \or %
         \let\next\spancount%
      \or %
         \let\next\argCOLskip%
      \else %
      \fi %
   \fi%
%  \out{subCOL-->  countREGISTER=[\the\countREGISTER]}
   \next%
}%
\gdef\argROWskip#1{%
%  Deletes the next balanced, undelimited argument from a
%                 token list.
% \out{---> Entering argROWskip <---}
% \out{In argROWskip:  deleted arg is [#1]}%
   \let\next\ROWcount \next%
}%  End of macro \argskip.
\gdef\arghdROWskip#1{%
%  Deletes the next balanced, undelimited argument from a
%                 token list.
% \out{---> Entering arghdROWskip <---}
% \out{In arghdROWskip:  deleted arg is [#1]}%
   \let\next\ROWcount \next%
}%  End of macro \arghdROWskip.
\gdef\argCOLskip#1{%
%  Deletes the next balanced, undelimited argument from a
%                 token list.
% \out{---> Entering argCOLskip <---}
% \out{In argCOLskip:  deleted arg is [#1]}%
   \let\next\COLcount \next%
}%  End of macro \argskip.
}%  End of active &'s.
}%  End of active |'s.
\def\spancount#1{%\out{spancount--->\meaning#1}
   \nspan=#1\multiply\nspan by 2\advance\nspan by -1%
   \global\advance \countREGISTER by \nspan
%  \out{number spancount--->\the\nspan; \the\countREGISTER}
   \let\next\COLcount \next}%
\def\dvr#1{\relax}%
% \omit\hfil%
% \parindent=0pt\hsize=1.1in\valign{%
% \vfil#\vfil&\vfil#\vfil\cr\hfil\hbox{\ Added to\ }\hfil&%
% \hfil\hbox{\ empty events\ }\hfil\cr}\hfil%
\def\header#1{%
\dvr{1}{\let\cr=\@mpersand%
\hdtks={#1}%
%\out{In header:  hdtks=[\the\hdtks]}%
\counthdROWS\hdtks\into\hdrows%
\advance\hdrows by 1%
\ifnum\hdrows=0 \hdrows=1 \fi%
%\out{In header:  Nhdrows=[\the\hdrows]}%
\dvr{5}\makehdPREAMBLE{\the\hdrows}%
%\out{In header:  headerpreamble=[\headerpreamble]}%
\dvr{6}\getHDdimen{#1}%
%\out{In header:  hdsize=[\the\hdsize]}%
%\striplastCR{#1}%
{\parindent=0pt\hsize=\hdsize{\let\ifmath0%
\xdef\next{\valign{\headerpreamble #1\crnorm}}}\dvr{7}\next\dvr{8}%
}%
}\dvr{2}}%  End of macro \header.
\def\makehdPREAMBLE#1{%This macro generates the necessary preamble for a
\dvr{3}%
%                      ruled table with \ncols primary columns.
%                      (Primary columns means the number of columns NOT
%                       counting those used for vertical rules.
\hdrows=#1%  Get the number of columns desired.
{%  Start local parameter definitions.
\let\headerARGS=0%
%  This is the key to the whole thing; it prevents \ARGS
\let\cr=\crnorm%
%                from being expanded in the followin \edef's.
\edef\xtp{\vfil\hfil\hbox{\headerARGS}\hfil\vfil}%
\advance\hdrows by -1%  One row has been generated; decrement the
%                         counter.
\loop%  Append as many further rows as needed to the preamble.
\ifnum\hdrows>0%
\advance\hdrows by -1%
\edef\xtp{\xtp&\vfil\hfil\hbox{\headerARGS}\hfil\vfil}%
\repeat%
\xdef\headerpreamble{\xtp\crcr}%
}%  End of local parameters.
\dvr{4}}%  End of \makehdPREAMBLE.
\def\getHDdimen#1{%
%\out{In getHDdimen:  Arg 1=[#1]}%
\hdsize=0pt%
\getsize#1\cr\end\cr%
}%  End of macro getHDdimen.
\def\getsize#1\cr{%
%\out{In getsize:  Arg 1=[#1]}%
%  Here we have to check arg#1 and see if the first token in #1 is an
%    \end; if so, we stop, else we check the width of arg#1.
%  We recall that each arg#1 will be terminated with a \cr token.
\endsizefalse\savetks={#1}%
%\out{In getsize:  the savetks = [\the\savetks]}%
\expandafter\lookend\the\savetks\cr%
%\out{In getsize:  ifendsize = [\meaning\ifendsize]}%
\relax \ifendsize \let\next\relax \else%
\setbox\hdbox=\hbox{#1}\newhdsize=1.0\wd\hdbox%
\ifdim\newhdsize>\hdsize \hdsize=\newhdsize \fi%
%\out{In getsize:  hdsize=[\the\hdsize]}%
%\out{In getsize:  newhdsize=[\the\newhdsize]}%
\let\next\getsize \fi%
\next%
}%
\def\lookend{\afterassignment\sublookend\let\looknext= }%
\def\sublookend{\relax%
%\out{In sublookend:  looknext = [\looknext]}%
\ifx\looknext\cr %
%\out{In sublooknext:  looknext=cr}%
\let\looknext\relax \else %
%\out{In sublooknext:  looknext/=cr}%
   \relax
   \ifx\looknext\end \global\endsizetrue \fi%
   \let\looknext=\lookend%
    \fi \looknext%
}%
%
%  Allow the user to make his own names for crthick, etc.
%
\def\tablelet#1{%
   \tableLETtokens=\expandafter{\the\tableLETtokens #1}%
}%
\catcode`\@=12%  Change @'s back to their normal category code.
%

% for numbering figures
\def\mvsmpi{1}		%m_vs_mpi.ps
\def\fourswaths{2}	%fourswaths.ps
\def\piswath{3}		%piswath.ps
\def\dmesxf{4}		%dmesxf.ps
\def\rrmesa{5}		%rrmesa.ps
\def\rrmesd{6}		%rrmesd.ps
\def\dbswaths{7}	%dbswaths.ps
\def\dnsusixswath{8}	%dnsusixswath.ps
\def\dnucx{9}		%dnucx.ps
\def\ddelx{10}		%ddelx.ps
\def\dbary{11}		%dbary.ps
\def\dneutcr{12}	%dneutcr.ps

% for numbering equations
\def\wvfn{1}
\def\cofrt{2}
\def\cexp{3}
\def\relwf{4}
\def\rsum{5}
\def\momentint{6}
\def\mesonwf{7}
\def\scalex{8}
\def\discretemoment{9}
\def\crsix{10}
\def\proposepsi{11}
\def\psiaab{12}
\def\psiaba{13}

% tables
\def\dmesxt{1}
\def\dxy{2}
\def\dnuccr{3}
\def\Tr{\rm Tr}
\if \preprint Y \twelvepoint\oneandathirdspace \fi
%\if \preprint N \twelvepoint\oneandathirdspace \fi	% *** kluge ****
 \preprintno{COLO-HEP-291}
\title Comparison of Lattice Coulomb Gauge Wave Functions in Quenched
Approximation and with Dynamical Fermions

\author
M.~W.~Hecht,${}^{(1)}$
Khalil M.~Bitar,${}^{(2)}$ T.~DeGrand,${}^{(3)}$ R.~Edwards,${}^{(2)}$
Steven Gottlieb,${}^{(4)}$ U.~M.~Heller,${}^{(2)}$ A.~D.~Kennedy,${}^{(2)}$
J.~B.~Kogut,${}^{(5)}$ W.~Liu,${}^{(6)}$
Michael C.~Ogilvie,${}^{(7)}$ R.~L.~Renken,${}^{(8)}$ Pietro~Rossi,${}^{(6)}$
D.~K.~Sinclair,${}^{(9)}$ R.~L.~Sugar,${}^{(10)}$
K.~C.~Wang${}^{(11)}$
\affil\vskip .10in
\centerline{${}^{(1)}$National Center for Atmospheric Research, P.O.
Box 3000, Boulder, CO 80307, USA}
\centerline{${}^{(2)}$SCRI, Florida State University, Tallahassee, FL
32306-4052, USA}
\centerline{${}^{(3)}$University of Colorado, Boulder, CO 80309, USA}
\centerline{${}^{(4)}$Indiana University, Bloomington, IN 47405, USA}
\centerline{${}^{(5)}$University of Illinois, Urbana, IL 61801, USA}
\centerline{${}^{(6)}$Thinking Machines Corporation, Cambridge, MA 02139, USA}
\centerline{${}^{(7)}$Washington University, St.~Louis, MO 63130, USA}
\centerline{${}^{(8)}$University of Central Florida, Orlando, FL 32816, USA}
\centerline{${}^{(9)}$Argonne National Laboratory, Argonne, IL 60439, USA}
\centerline{${}^{(10)}$University of California, Santa Barbara, CA 93106, USA}
\centerline{${}^{(11)}$University of New South Wales, Kensington, NSW 2203,
Australia}
\goodbreak
%\if \preprint Y \preprintno{COLO-HEP-291}\fi
\abstract

We present a comparison of Coulomb gauge wave functions from
$6/{g^2}=6.0$ quenched simulations with two simulations which include
the effects of dynamical fermions: simulations with two flavors of
dynamical staggered quarks and valence Wilson quarks at $6/{g^2}=5.6$
and simulations with two flavors of dynamical Wilson quarks and Wilson
valence quarks, at $6/{g^2}=5.3$.  The spectroscopy of these systems
is essentially identical. Parameterizations of the wave functions are
presented which can be used as interpolating fields for spectroscopy
calculations. The sizes of particles are calculated using these
parameterized wave functions.  The resulting sizes are small,
approximately half the sizes of the physical states.  The charge
radius of the neutron, which provides an indication of the asymmetries
between the wave functions of up and down quarks, is calculated.
Although the size of the nucleon in these simulations is small, the
ratio of the charge radius of the neutron to that of the proton is
consistent with the physical value. We find no significant differences
between the quenched and dynamical simulations.
\endpage
\body
\head{I. Introduction}
Numerical studies of QCD have become sufficiently fine-grained that it
has become possible to investigate the global properties of QCD wave
functions directly from Monte Carlo simulations.  The goal of these
studies is two-fold: First, visualizing wave functions is a powerful
diagnostic for lattice studies.  A picture of the wave function
provides a hint for a good trial wave function for spectroscopy.  One
can see whether the wave function of a hadron is squeezed by the
simulation volume; if it is, then a calculation of spectroscopy may be
compromised.

Second, it may be possible to use wave functions for phenomenology,
either by directly performing calculations with the wave functions, or
by abstracting a continuum model from the wave function, determining
its parameters from a small number of lattice measurements, and using
the model, rather than expensive lattice simulations, for QCD
calculations. Phenomenologically interesting calculations include
charge radii and radial moments. Results to date indicate that the
wave functions for hadrons in quenched QCD are too small in spatial
extent to reproduce quark phenomenology, although ratios of sizes,
including the ratio of the charge radius of the neutron to that of the
proton, are reasonable\rlap.\refto{SWAVE}

The subject goes back for many years.  In 1985 Velikson and
Weingarten\refto{VELIK} studied meson wave functions in SU(2) and
Gottlieb\refto{GOTT} carried out the first study of wave functions
with SU(3) Wilson fermions.  Recently, Chu, Lissia and Negele have
investigated gauge invariant wave functions (with a product of links
connecting the quarks)\rlap.\refto{NEGELE} Wave functions for heavy
quark systems\refto{FNAL} and for heavy-light systems\refto{HL} have
also recently been reconstructed, and two of us\refto{{PWAVE},{SWAVE}} have
performed an extensive study of wave functions of light quark systems.
Wave function methods have also been applied to finite temperature
systems, in Ref.~\cite{MILC}.

In this paper we extend the wave function calculations of Ref.
\cite{SWAVE} to systems with dynamical fermions, using lattices
generated as part of the High Energy Monte Carlo Grand Challenge.  We
parameterize the wave functions for possible use as interpolating
fields for spectroscopy.  In addition, we compare the charge radii and
radial moments determined from the wave functions to the
experimentally determined numbers and to the values obtained in the
quenched approximation\rlap.\refto{SWAVE} We note that these wave
functions are minimal Fock space wave functions and that the use of
the wave function for calculating phenomenological numbers represents
an uncontrolled approximation.

The wave function $\psi_G(r)$ of a meson H in a gauge G is defined as
\ $$\psi_G(r) =
\sum_{\vec x} \langle H | q(\vec x)
{\bar q}(\vec x + \vec r) | 0 \rangle\eqno(\wvfn)$$
where  $q(\vec x)$ and ${\bar q}(\vec y)$  are
quantum mechanical operators which create a quark and an antiquark at
locations $\vec x$ and $\vec y$. (We have suppressed Dirac and color
indices.)  The wave function can be extracted from a correlation
function which is a convolution of quark and antiquark propagators
$G(x,y)$
\ $$C(\vec r,t) = \sum_{\vec x}\Psi(\vec y_1, \vec y_2)
G_q(\vec y_1,0;\vec x,t)
G_{\bar q}(\vec y_2,0;\vec x + \vec r,t)
\eqno(\cofrt)$$
where $\Psi(\vec y_1, \vec y_2)$ is the $t=0$ operator.
At large $t$ if the mass of the hadron is $m_H$, then
\ $$C(\vec r,t ) \simeq \exp (-m_H t) \psi_G(\vec r)
\eqno(\cexp)$$
and so by plotting $C(\vec r, t)$ as a function of $\vec r$ we can
reconstruct the wave function up to an overall constant.  One can
derive a similar expression for baryons, as a function of the two
relative coordinates of the three valence quarks.

\head{II. The simulations}
Our simulations were performed on Connection Machine CM-2s located at
the Supercomputer Computations Research Institute at Florida State
University and at the Pittsburgh Supercomputing Center.

The quenched data set consists of 41 lattices of size $16^4$ sites at
a coupling $6/{g^2}=6.0$ separated by 500 evolutionary sweeps (100
passes through the lattice of a pattern of four overrelaxed
sweeps\refto{RELAX} and one Kennedy-Pendleton heat bath
sweep\refto{KPQ}).  We recorded propagators with hopping parameters
equal to $\kappa=0.145,0.152,0.153,0.154,0.155$ with the corresponding
pion masses in lattice units range from $m_{\pi}a=0.82$ to
$m_{\pi}a=0.28$, where $a$ is the lattice spacing.

The simulations with two flavors of dynamical staggered quarks use the
Hybrid Molecular Dynamics algorithm\rlap.\refto{HMD} The lattice size
is $16^3 \times 32$ sites and the lattice coupling $6/{g^2}=5.6$.  The
dynamical quark mass is $am_q=0.01$.  A subset of the data (whose
spectroscopic analysis is described in Ref.~\cite{LATESTGC}) was taken
for this analysis. It consists of 20 lattices spaced 80 simulation
time units apart (with the normalization of Ref.~\cite{HEMCGC}). We
computed spectroscopy with staggered sea quarks at three values of the
Wilson quark hopping parameter: $\kappa=0.1565$, 0.1585, and 0.1600.
The pseudoscalar mass in lattice units ranges from about 0.22 to 0.45.

The simulations with two flavors of Wilson sea quarks used the Hybrid
Monte Carlo algorithm\rlap.\refto{HMC} The lattice size is again $16^3
\times 32$ and the lattice coupling is $6/{g^2}=5.3$. Again, a subset of
the whole data set (whose spectroscopic analysis will be described in
Ref.~\cite{GCW}) was taken consisting of 19 lattices spaced 65 Hybrid
Monte Carlo time units apart. Only one hopping parameter was studied:
$\kappa=0.1670$, corresponding to a pion mass in lattice units of
about 0.47.

All spectroscopy in the three data sets was extracted using identical
methods and computer programs.  We gauge fixed lattices to Coulomb
gauge using an overrelaxation algorithm\rlap.\refto{OVERRELAX} Our
criterion for gauge fixing was that the average change in the trace of
a spacelike link was less than $\Tr \delta U = 10^{-5}$.  The sources
for the quarks are Gaussians centered about some origin on a single
time slice.  Our inversion technique is conjugate gradient with
preconditioning via ILU decomposition by
checkerboards\rlap.\refto{DEGRANDILU} We used a fast matrix inverter
written in  CMIS (Connection Machine Instruction Set)\refto{LIU}.

We employ relativistic wave functions.\refto{WILSONWAVES} The baryon
wave functions are:
\ $$
\eqalign{
\noalign{\hbox{Proton:}}
 \left|P,s\right>& = (u C\gamma_5 d)u_s \cr
                 &=(u_1 d_2 - u_2 d_1 + u_3 d_4 -u_4 d_3) u_s\quad
\cr
\noalign{\hbox{Delta:\ }}
\left|\Delta,s\right>&= (u_1 d_2 + u_2 d_1 + u_3 d_4 + u_4 d_3) u_s\quad
\cr}
\eqno{(\relwf)}$$
We measured meson correlation functions using spin structures for the
source of $\bar \psi \gamma_5 \psi$ for the pion and $\bar \psi
\gamma_0\gamma_3 \psi$ for the rho. At the wave function we used the
same spin structure for the pion and $\bar \psi \gamma_3 \psi$ for the
rho.

We include the full covariance matrix in order to get a meaningful
estimate of the goodness of fit.  Reference \cite{DOITRIGHT} discusses
this fitting procedure in detail.

The rho mass was used to fix the spacing on the dynamic staggered
lattices at $a^{-1}=2140{\rm\ MeV}$\rlap,\refto{LATESTGC} which
differs only slightly from the lattice spacing in quenched QCD at
$6/g^2=6.0$ of $a^{-1}=2312{\rm\ MeV}$\rlap.\refto{APE} Comparison of
the two numbers suggests that the spacing on the dynamic staggered
lattices is 8\% larger than on the quenched lattices.  Fixing the
lattice spacing to the proton mass yields a value of $a^{-1}=1800
{\rm\ MeV}$ on the dynamic staggered fermion lattices\refto{LATESTGC}
and $a^{-1}=1991 {\rm\ MeV}$ on the quenched
lattices\rlap.\refto{PMASS} We use $a^{-1}=2000 {\rm\ MeV}$ to
estimate dimensionful quantities for both the dynamical staggered
fermion data and the quenched simulation.

We have recently extended the dynamic Wilson spectroscopy to a second
value of the Wilson hopping parameter at
$\kappa=0.1675$\rlap.\refto{GCW} The rho mass fixes the lattice spacing
to $a^{-1}=1640{\rm\ MeV}$, indicating that the spacing of the dynamic
Wilson lattice is around 30\% larger than the spacing of the dynamic
staggered lattice.  As a lattice problem we can analyze the wave
functions on the dynamical Wilson fermion lattices and compare their
properties (such as their sizes) to those in the quenched and
staggered dynamical simulations when the lattice masses are similar,
providing another comparison of the lattice spacings.  Indeed, we have
a similar problem comparing the quenched and staggered dynamical
fermion simulations: all the bare parameters are different.  However,
when we compare mass ratios (via Edinburgh plots, for example), we see
that the data sets are not dissimilar.

The most striking way we have found to display spectroscopy from the
three data sets is to plot the vector and baryon masses as a function
of the pion mass in lattice units. This we do in Fig.~\mvsmpi.  We see that
the the three data sets resemble each other rather closely, though the
Wilson dynamical fermion particles appear to be about fifteen per cent
heavier than the quenched and staggered dynamical spectroscopy at the
same value of the pion mass.

\if\preprint Y \psfigure 252 108  {Figure~\mvsmpi} {m_vs_mpi.ps}
{Comparison of quenched Wilson spectroscopy at $6/{g^2}=6.0$ (squares)
with Wilson valence spectroscopy from a dynamical staggered fermion
simulation at at $6/{g^2}=5.6$ (octagons) and dynamical Wilson fermion
simulation at $6/{g^2}=5.3$ (diamond).}
\fi

\head{III. Global Views of Wave Functions}

A hierarchy of particle sizes emerges from a comparison of the wave
functions. To facilitate this comparison the meson wave functions
plotted in Fig.~\fourswaths\ have been normalized so that the value at
zero separation is one.  The baryon wave functions show greater
fluctuation in the normalization than the meson wave functions. We
have normalized the baryon wave functions in Fig.~\fourswaths\ on a
lattice-by-lattice basis.  Our justification for presenting baryons in
this way is that the resulting plot is consistent with that obtained
from a correlated fit to the data, and doing so helps the viewer to
see qualitative features.  The meson wave functions show the amplitude
as the quark is pulled apart from the antiquark along a principal
axis.  The baryon plots are of the wave functions for the unique
flavor quark when the two like-flavor quarks are fixed to be at the
same site.  The pion and proton wave functions are smallest; the rho
is largest, and the delta is next largest.

\if\preprint Y \psfigure 468 0  {Figure~\fourswaths} {fourswaths.ps}
{Coulomb gauge wave functions at time $t=6$, for separations of
$(x,0,0)$.  For the baryons the two like-flavor quarks are pinned to
the same site while the non-like quark is at separation $(x,0,0)$
relative to the other two. The meson data has been normalized after
averaging. The baryon data has been normalized on a lattice-by-lattice
basis.  Particles are labeled by boxes for pion, diamonds for rho,
octagons for proton and crosses for delta.  (a)---(c) are simulations
with dynamical staggered quarks and Wilson valence quarks, at Wilson
hopping parameters of $\kappa=0.1565, 0.1585$ and $0.1600$,
respectively. (d) is with Wilson dynamical and valence quarks at
$\kappa=0.1670$.  }
\fi

The wave functions for the hadrons made of the lightest valence quarks
are very large.  Of course, because of the periodic boundary
conditions in the spatial directions of the lattice, the wave
functions shown at $r=8$ are twice the size they would be on an
infinite size lattice. Nevertheless, the $\kappa=0.1600$ rho has only
fallen to twenty five per cent of its peak value by $r=8$.

The pion wave functions in the staggered and Wilson simulations are
compared in Fig.~\piswath. The pion wave function on the staggered
lattices is relatively insensitive to the value of the quark mass. The
pion wave function on the Wilson lattices is dramatically smaller,
measured in lattice units, although the masses in lattice units on the
Wilson and the $\kappa=0.1565$ staggered dynamic fermion lattices are
similar.  Our wave function analysis supports the 30\% difference in
the lattice spacings indicated by fixing the lattice spacings to the
rho mass, as shown in the next sections.

\if\preprint Y \psfigure 252 108  {Figure~\piswath} {piswath.ps}
{Pion wave function on lattices with staggered fermions at
all three $\kappa$ values ($\kappa=0.1565,0.1585,0.1600$) and with
Wilson fermions (at $\kappa=0.1670$).}
\fi

All of the spectroscopy with dynamical staggered fermions was
originally performed on spatial lattices with $12^3$ sites.  Both
baryons showed strong finite-size effects: their masses fell by about
fifteen per cent when they were recomputed on a $16^3$ lattice.
Neither meson showed an appreciable change in mass with lattice size.
It is difficult to reconcile this behavior with the observed hierarchy
of wave functions: why are finite size effects not largest for the rho
meson?  Some physics which governs the energy of a particle in a
finite simulation volume is not being included in the minimum Fock
space wave function.

\head{IV. Comparisons of Quenched and Dynamical Simulations}
\subhead{A. Meson Properties}
In this and the following section we analyze the wave functions.  We
parameterize the wave functions for possible use as interpolating
fields for spectroscopy.  In addition, we compare the charge radii and
radial moments determined from the wave functions to the
experimentally determined numbers and to the values obtained in the
quenched approximation. All of the data in quenched approximation has
been presented and more completely discussed in Ref.~\cite{SWAVE}.  We
remind the reader that these wave functions are minimal Fock space
wave functions and that the use the wave function for calculating
phenomenological numbers represents an uncontrolled approximation.

The second moment of the pion, $\langle r^2_\pi\rangle$, has been
determined experimentally to have the value of $0.405\pm0.024 $
fm${}^2$.\refto{DALLY} The second moment is defined in the quark model
as
\ $$
\langle r^2\rangle=\sum_i q_i\langle(\vec r_i - \vec R)^2\rangle
\eqno{(\rsum)} $$
where $\vec R$ is the location of the center of mass and $q_i$ is the
charge of the $i^{th}$ valence quark. In terms of the wave function
$\Psi$ this is
\ $$
\langle r^2\rangle = {\int d^3{\vec x}\ ({x \over 2})^2 \Psi^2({\vec
x}) \over {\int d^3{\vec x}\ \Psi^2({\vec x})}}.
\eqno{(\momentint)} $$

In order to evaluate the second moment from our data we suggest a
parameterization of the wave function, make a correlated fit of the
parameters to the data, and integrate analytically to obtain the
second moment at each value of the hopping parameter.

Our parametrization of the meson wave function is
\ $$
\Psi(r)=x_1 \exp(-x_2 r^{x_3})
\eqno{(\mesonwf)} $$
where $r$ is the separation between quark and antiquark.  The periodic
boundary conditions are treated by including an additional term with
$(L-r)$ substituted in place of $r$, where $L$ is the length of the
lattice, as in Ref.~\cite{SWAVE}.  A full correlated fit of these
three parameters to a subset of the data is made. We choose to use the
points along principal axes of the lattice in the fit.  The resulting
fit parameters are listed in Table~\dmesxt\ and are plotted in
Fig.~\dmesxf.

\if\preprint Y \psoddfigure  468 226 0  {Figure~\dmesxf} {dmesxf.ps}
{Correlated fit parameters for meson wave
functions (as in Eqn.~\mesonwf) from correlated fits to the data. (a)
exponential falloff $x_2$, (b) exponent $x_3$.}
\fi

\if\preprint Y
\vbox{
\bigskip
\centerline{ TABLE~\dmesxt: Correlated Fit Parameters for Mesons}
\bigskip
\begintable
$\kappa$~|sea quarks~|pion $x_2$~|pion $x_3$~|rho $x_2$~|rho $x_3$\cr
0.145~|quenched~|0.2069(16)~|1.274(9)~|0.0972(14)~|1.514(11)~\cr
0.152~|quenched~|0.1970(10)~|1.247(7)~|0.0727(12)~|1.534(15)~\cr
0.153~|quenched~|0.1964(9)~|1.241(7)~|0.0698(14)~|1.534(17)~\cr
0.154~|quenched~|0.1961(9)~|1.234(7)~|0.0674(18)~|1.532(21)~\cr
0.155~|quenched~|0.1960(10)~|1.228(8)~|0.0658(26)~|1.529(28)~\cr
0.1565~|staggered~|0.234(2)~|1.216(7)~|0.094(3)~|1.45(2)~\cr
0.1585~|staggered~|0.232(2)~|1.211(8)~|0.086(4)~|1.46(3)~\cr
0.1600~|staggered~|0.234(2)~|1.213(11)~|0.080(5)~|1.48(3)~\cr
0.1670~|Wilson~|0.337(5)~|1.253(9)~|0.121(4)~|1.58(2)
\endtable
}
\fi

The exponent $x_3$ is close to 3/2 for the rho meson wave function
calculated with dynamic fermions, as it is for the quenched rho.  This
is the value obtained as the solution to the nonrelativistic wave
equation in a linear potential, and thus may be an indication of a
potential which is approximately linear in the quark separation.

The second moments of the mesons calculated from correlated fit
parameters on both staggered and Wilson lattices are shown in
Fig.~\rrmesa. The mass of the dynamic staggered pion, at
$am_\pi=0.45$, is comparable to the mass of the dynamic Wilson pion,
at $am_\pi=0.47$. The second moments of the dynamic staggered mesons
are approximately twice as large as the second moments of the dynamic
Wilson mesons; the ratios for both pion and rho meson are $2.1\pm0.1$.
This difference is largely explained by the 30\% difference in lattice
spacings found by fixing the lattice spacings to the rho mass.

\if\preprint Y \psoddfigure  468 226 0  {Figure~\rrmesa} {rrmesa.ps}
{Second moment of mesons (as in Eqn.~\momentint), using
parameterized wave functions of Eqn.~\mesonwf.  Crosses indicate
staggered dynamic fermions, squares indicate Wilson dynamic fermions,
diamonds are quenched.  (a) pion, (b) rho.  }
\fi

The correlated fit parameters scale in a way which is roughly
consistent with a 30\% difference in the lattice spacings. The value
of the exponent $x_3$ is unchanged by a rescaling of the lattice
spacing, but the exponential falloff $x_2$ is rescaled as
\ $$
x_2'=(a'/a)^{x_3}x_2.
\eqno(\scalex)$$
In fact the values of $x_3$ are not very different in the dynamic
staggered and dynamic Wilson simulations. We would expect the ratio of
$x_2$ for the Wilson point to the $\kappa=0.1565$ staggered point to
be around $1.38$ for the pion and around $1.49$ for the rho meson
based on the scaling relation of Eqn.~\scalex. From Table~\dmesxt\ we
find the ratio of exponential falloffs to be $1.44\pm0.02$ for the
pion, which is a bit higher than the anticipated value and would
suggest a 40\% difference in the lattice spacings. For the rho meson
the ratio is $1.29\pm0.06$, which is lower than the anticipated value
and would suggest a 20\% difference in the lattice spacings.

Extrapolating the second moment of the dynamic staggered pion linearly
in $\kappa$ to $\kappa_c$, we find
$\langle(r/a)_\pi^2\rangle=4.79\pm0.20$, a figure which is three
standard deviations below the quenched value of $6.24\pm0.25$.  The
corresponding number for the dynamic staggered rho is
$\langle(r/a)_\rho^2\rangle=8.91\pm0.72$, which is around one standard
deviation below the quenched value of $10.3\pm1.1$.  The ratio of the
moments is $\langle r_\pi^2\rangle/\langle
r_\rho^2\rangle=1.86\pm0.17$, which is consistent with the quenched
ratio of $1.65\pm0.19$.  The quenched pion second moment is 30\% larger
than that of the dynamic pion.  This difference in size could be
completely explained by a 15\% difference in lattice spacings, and
could be at least partially explained by the approximately 8\%
difference in the quenched and dynamic staggered lattice spacings
determined through fixing the lattice spacing to the mass of the rho
meson.  The second moment of the quenched rho meson is 15\% larger
than that of the dynamic rho, a difference which is perfectly
accounted for by an 8\% difference in the lattice spacings.

The pion moment on the staggered fermion lattices converts
approximately to the dimensionful number of $\sqrt{\langle
r^2_\pi\rangle}=0.21{\rm \ fm}$, which is one third of the physical
value, and for the rho meson $\sqrt{\langle r^2_\rho\rangle}=0.29{\rm
{}~fm}$, using $a^{-1}=2000{\rm\ MeV}$.

The only qualitative feature we observe for the dynamic mesons which
may differ from the quenched mesons is the dependence of the pion size
on the value of the hopping parameter.  The quenched pion's size was
observed to grow consistently larger with decreasing quark mass.  The
dynamic pion at $\kappa=0.1600$ appears to be no larger than the
dynamic pions at the two smaller $\kappa$ values, as is seen in
Fig.~\rrmesa.

The second moments of the mesons can be calculated directly from the
data using discrete lattice sums, as
\ $$
<r^2>=\bigl(\sum_{lattices}\sum_{\vec s}(s/2)^2 \Psi^*(\vec s)
\Psi(\vec s)\bigr)/ \bigl(\sum_{lattices}\sum_{\vec s}\Psi^*(\vec s)
\Psi(\vec s)\bigr).
\eqno{(\discretemoment)}$$
These second moments calculated from discrete lattice sums are shown
in Fig.~\rrmesd.  Using this method to compute the second moments of
mesons on the staggered lattices at all three $\kappa$ values, a
linear extrapolation in $\kappa$ to $\kappa_c$ results in a pion
second moment of $\langle (r/a)_\pi^2\rangle=6.32\pm0.58$, one and a
half standard deviations below the corresponding quenched number of
$8.05\pm0.65$.  The second moment of the rho meson is found to be
$\langle (r/a)_\rho^2\rangle=13.7\pm0.6$, which is consistent with the
quenched number of $14.0\pm0.8$.  In contrast to the suggestion that
the second moment of the pion as derived from correlated fits is
independent of quark mass, the second moment from the discrete lattice
sums rises steadily with decreasing quark mass.  This
method of obtaining radial moments does not compensate for the
contributions to the wave function from image particles, nor does it
account for the considerable tails of the wave functions which extend
beyond the lattice. Radial moments derived from correlated fits to
the data, as outlined near the beginning of this section, are free of
these two problems.

\if\preprint Y \psoddfigure  468 226 0  {Figure~\rrmesd} {rrmesd.ps}
{Meson second moments as calculated through discrete lattice
sums, via Eqn.~\discretemoment.  Crosses indicate staggered dynamic
fermions, squares indicate Wilson dynamic fermions, diamonds are
quenched. (a) pion, (b) rho.}
\fi

The second moments of the mesons on the staggered lattices are slightly
smaller than the second moments which were calculated in the quenched
approximation.  The ratios of the second moments of rho meson to pion
are completely consistent between the two simulations.  It is possible
that all of the difference in the sizes can be ascribed to a difference
in the lattice spacings.  As observed for the quenched pion, the pion on
a lattice containing dynamic staggered fermions has a size which is
approximately one third that of the physical pion.

\subhead{B. Baryon Properties}

Charge radii for the baryons cannot be calculated by discrete lattice
sum owing to the limited subset of the data which has been recorded,
but the charge radii can be calculated by parametrizing the wave
function, inserting the resulting expression for the wave function in
the analytic expression for the charge radius and integrating. We do
this to compare the charge radii of the baryons on dynamic
lattices with charge radii in the quenched approximation, as well as
for comparison with the physical values.

The integral for the charge radius of a baryon is written in terms of
two relative coordinates, one of which is the separation between two
quarks of flavor $a$ ($\vec r_{aa}$), and the second of which is a
vector reaching from midway between the $a$ quarks to a quark of
flavor $b$ ($\vec r_{cb} = \vec r_b - {1 \over 2}(\vec r_a + \vec r_{a'})$).
In terms of these variables the integral for the charge radius is
\ $$
\langle r^2\rangle = {\int d^3\vec r_{aa}\int d^3\vec r_{cb} |\Psi(\vec
r_{aa},\vec r_{cb})|^2 \sum_{q=1,2,3} e_q r_q^2(\vec
r_{aa},\vec r_{cb}) \over \int d^3\vec r_{aa}\int d^3\vec r_{cb}
|\Psi(\vec r_{aa},\vec r_{cb})|^2}
\eqno{(\crsix)} $$
where $r_q$ is the distance from the center of mass to the location of
a particular quark.

We compare the probability for the two quarks of flavor $a$ to be at
the same position, with the quark of flavor $b$ out at some distance
$r$. If SU(6) were unbroken then this would be equal to the
probability for one of the $a$ quarks to be at the same position as
the $b$ quark, with the second $a$ quark out at the same distance $r$.
Swaths of such points are compared in Fig.~\dbswaths. The nucleon wave
function amplitudes at separations 2, 4 and 6 for the two orientations
differ by about three standard deviations, indicating a negative
charge radius for the neutron. The delta wave function amplitudes for
the two orientations differ by about ${3\over2}\sigma$, indicating a
slight positive charge radius for the delta of quark content $ddu$.
This is unanticipated and may represent a statistical fluctuation. We
note again that the baryon wave function points are normalized on a
lattice-by-lattice basis.

\if\preprint Y \psoddfigure  468 226 0  {Figure~\dbswaths} {dbswaths.ps}
{Falloff of wave function of $b$ quark (crosses) with
separation from an $aa$ diquark, and of an $a$ quark from an $ab$
diquark (octagons). (a) nucleon, (b) delta. Data are at
$\kappa=0.1585$, with dynamic staggered fermions.}
\fi

We use a wave function which is a product of three exponentials, with
each exponential being a function of the separation between one of the
pairs of quarks. Our wave function is, for two quarks of flavor
``$a$'' and one quark of flavor ``$b$''
\ $$
\Psi(r_{aa},r_1,r_2)=N\exp(-x_{aa} r_{aa}^y)
\exp(-x_{ab} r_1^y)
\exp(-x_{ab}r_2^y)
\eqno{(\proposepsi)} $$
where $r_{aa}$ is the relative separation of the two $a$ quarks (as in
Eqn.~\crsix), $r_1$ is the separation of the $b$ quark from an $a$
quark and $r_2$ is the separation between the $b$ quark and the other
$a$ quark.

We store data at four separations between the like-flavor quarks
(quarks of flavor $a$ in Eqn.~\proposepsi), and for each of those four
values we store the amplitude for the other quark (flavor $b$ in
Eqn.~\proposepsi) to be anywhere on the lattice.  In order to
calculate charge radii we make use of a very limited but symmetric
subset of the data, using the four data points for which the two
quarks of flavor $a$ are at zero relative separation and the quark of
flavor $b$ is at separations of 0, 2, 4 and 6 from the $aa$ pair, as
well as the points for which one $a$ quark is at zero relative
separation from the $b$ quark and the second quark of flavor $a$ is at
separations of 2, 4 and 6 from the $ab$ pair.

For this subset of the data the parametrized wave function of
Eqn.~\proposepsi\ can be simplified.  The wave function for a quark of
flavor $b$ relative to an $aa$ diquark can be written as
\ $$
\Psi_b(r_b) = N \exp(-x_br_b^y).
\eqno{(\psiaab)} $$
and the wave function for a quark of flavor $a$ relative to an
$ab$ diquark is written as
\ $$\Psi_a(r_a) = N \exp(-x_ar_a^y).
\eqno{(\psiaba)} $$
Full correlated fits are made of the parameters of Eqns.~\psiaab\ and
\psiaba\ to this limited subset of the data. The data points and the
functional form which has been fit to those points are illustrated in
Fig.~\dnsusixswath\ for the $\kappa=0.1585$ nucleon.

\if\preprint Y \psfigure 200 134  {Figure~\dnsusixswath} {dnsusixswath.ps}
{Nucleon wave function data points at $\kappa=0.1585$ (with dynamic
staggered fermions) with functional forms of Eqn.~\psiaab\ and
Eqn.~\psiaba\ overplotted.  Crosses represent wave function for quark
of flavor $b$ relative to $aa$ diquark, octagons represent wave
function for quark of flavor $a$ relative to $ab$ diquark.}
\fi

Baryon correlated fit parameters are presented in Table~\dxy\ and in
Figs.~\dnucx, \ddelx\ and \dbary. The value of the exponent $y$ lies
in a narrow range for all of the baryons, regardless of the
composition of the lattice or of the identity of the baryon. The
magnitude of $y$ varies only slightly with quark mass. The exponential
falloffs $x_{aa}$ and $x_{ab}$ tend towards slightly larger values on
the dynamic staggered lattices than on the quenched lattices,
similarly suggesting a small difference in the lattice spacings in the
two formulations. The magnitude of the exponential falloffs between a
pair of $aa$ quarks and an $ab$ pair is undifferentiated for the delta
within each of the four dynamic fermion simulations.  This indicates a
neutral charge radius for the delta, despite the hint of a positive
charge radius from the wave function points of Fig.~\dbswaths. The
magnitude of the exponential falloffs between one pairing of the
quarks is substantially different from that between the other pairing
of quarks for the nucleon within each simulation, indicating a
statistically significant negative charge radius for the neutron.

\if\preprint Y \psfigure 200 134  {Figure~\dnucx} {dnucx.ps}
{Exponential falloffs $x_{aa}$ and $x_{ab}$ which
parameterize the nucleon wave function, as in Eqn.~\proposepsi. }
\fi
\if\preprint Y \psfigure 200 134  {Figure~\ddelx} {ddelx.ps}
{Exponential falloffs $x_{aa}$ and $x_{ab}$ which
parameterize the wave function of the delta, as in Eqn.~\proposepsi. }
\fi
\if\preprint Y \psoddfigure  414 200 27  {Figure~\dbary} {dbary.ps}
{Value of the exponent $y$ which parametrizes the baryon
wave functions, as in Eqn.~\proposepsi. (a) nucleon, (b) delta. }
\fi

\if\preprint Y
\vbox{
\bigskip
\centerline{ TABLE~\dxy: Correlated Fit Function Parameters for
Baryons }
\centerline{(a): Nucleon}
\bigskip
\begintable
$\kappa$~|sea quarks~|$x_a$~|$x_b$~|$x_{aa}$~|$x_{ab}$~|y\cr
0.145~|quenched~|0.158(3)~|0.177(3)~|0.069(3)~|0.089(1)~|1.398(12)\cr
0.152~|quenched~|0.135(3)~|0.156(2)~|0.057(3)~|0.078(1)~|1.378(11)\cr
0.153~|quenched~|0.132(3)~|0.153(2)~|0.055(3)~|0.077(1)~|1.372(12)\cr
0.154~|quenched~|0.129(3)~|0.151(3)~|0.054(4)~|0.076(1)~|1.365(14)\cr
0.155~|quenched~|0.125(4)~|0.149(3)~|0.050(5)~|0.074(2)~|1.361(19)\cr
0.1565~|staggered~|0.158(4)~|0.187(4)~|0.065(4)~|0.094(2)~|1.363(12)\cr
0.1585~|staggered~|0.148(4)~|0.180(4)~|0.058(5)~|0.090(2)~|1.355(12)\cr
0.1600~|staggered~|0.123(10)~|0.170(5)~|0.038(10)~|0.085(3)~|1.35(4)\cr
0.1670~|Wilson~|0.208(9)~|0.264(8)~|0.076(10)~|0.132(4)~|1.33(3)
\endtable
\bigskip
\centerline{(b): Delta}
\bigskip
\begintable
$\kappa$~|$x_a$~|$x_b$~|$x_{aa}$~|$x_{ab}$~|y\cr
0.145~|0.150(3)~|0.141(3)~|0.079(3)~|0.071(1)~|1.410(13)\cr
0.152~|0.124(3)~|0.116(2)~|0.066(3)~|0.058(1)~|1.376(15)\cr
0.153~|0.122(3)~|0.113(2)~|0.065(4)~|0.057(1)~|1.367(17)\cr
0.154~|0.120(4)~|0.111(2)~|0.064(4)~|0.056(1)~|1.357(19)\cr
0.155~|0.119(6)~|0.109(3)~|0.064(6)~|0.055(2)~|1.346(26)\cr
0.1565~|0.159(6)~|0.155(5)~|0.081(6)~|0.078(3)~|1.323(23)\cr
0.1585~|0.152(8)~|0.154(6)~|0.075(9)~|0.077(3)~|1.31(3)\cr
0.1600~|0.159(20)~|0.142(9)~|0.088(21)~|0.071(4)~|1.26(6)\cr
0.1670~|0.217(17)~|0.226(26)~|0.104(22)~|0.113(13)~|1.28(11)
\endtable}
\fi

The baryon correlated fit parameters are consistent with a 30\%
difference in the lattice spacings.  For a 30\% difference in
lattice spacings we expect the ratio of exponential falloffs on the
dynamic Wilson lattices to that on the dynamic staggered lattices at
$\kappa=0.1565$ to be around $1.42$ for the nucleon, based on the
scaling relation of Eqn.~\scalex. In fact we find the ratio of
$x_{aa}$ parameters to be $1.18\pm0.18$, and the ratio for the
parameter $x_{ab}$ is $1.41\pm0.06$. Both of these figures compare
well with the expected value of $1.42$. For the delta the anticipated
ratio of exponential falloffs is $1.41$. We find the ratio of
$x_{aa}$'s for the delta is $1.3\pm0.3$, and the ratio of $x_{ab}$'s
is $1.4\pm0.2$, both of which are consistent with the expected ratio.

The charge radii for the proton and neutron are presented in
Table~\dnuccr.  A linear extrapolation in $\kappa$ to $\kappa_c$ of
the proton charge radius yields a value of $\langle(r/a)^2_p\rangle=
15.1\pm3.4$, slightly below but consistent with the charge radius in
the quenched approximation of $16.6\pm3.1$.  The ratio of the charge
radius of the neutron to that of the proton is in Table~\dnuccr\ and
is also displayed in Fig.~\dneutcr.  The dynamic data are consistent
with the quenched data except at $\kappa=0.1600$, where the point
falls one standard deviation low.  A linear extrapolation in $\kappa$
of the ratio of the charge radii on the dynamic staggered lattices
gives $\langle r^2_n\rangle/\langle r^2_p\rangle=-0.21\pm0.04$.
This ratio is slightly more than one standard deviation above the
experimental figure of $-0.146\pm0.005$\rlap,\refto{CHARGERATIO} and is
evidently pulled in that direction by the point at $\kappa=0.1600$.
The charge radius of the proton calculated with dynamic staggered
fermions is $\sqrt{\langle r^2_p\rangle}=0.38{\rm\ fm}$ using
$a^{-1}=2000{\rm\ MeV}$, half the physical size of
$\sqrt{\langle r^2_p\rangle}=0.81$ fm\rlap.\refto{CRPROTON}

\if\preprint Y \psfigure 252 108  {Figure~\dneutcr} {dneutcr.ps}
{Ratio of charge radii of neutron to proton as a function of
pion mass. Horizontal dashed line is the experimental ratio. Crosses
indicate staggered dynamic fermions, squares indicate Wilson dynamic
fermions, diamonds are quenched. }
\fi

\if\preprint Y
\vbox{
\bigskip
\centerline{ TABLE~\dnuccr: Charge Radii of Baryons}
\bigskip
\begintable
$\kappa$~|sea quarks~|$\langle (r/a)^2_n\rangle$~|
$\langle (r/a)^2_p\rangle$~|
$\langle r^2_n\rangle/\langle r^2_p\rangle$\cr
0.145~|quenched~|-0.73(38)~|9.1(12)~|-0.080(34)\cr
0.152~|quenched~|-1.27(59)~|12.4(18)~|-0.102(37)\cr
0.153~|quenched~|-1.37(64)~|13.1(20)~|-0.104(38)\cr
0.154~|quenched~|-1.55(71)~|14.0(23)~|-0.111(38)\cr
0.155~|quenched~|-1.83(82)~|15.0(29)~|-0.122(38)\cr
0.1565~|staggered~|-1.23(33)~|10.5(12)~|-0.117(23)\cr
0.1585~|staggered~|-1.63(43)~|12.1(15)~|-0.135(24)\cr
0.1600~|staggered~|-3.5(18)~|17.4(61)~|-0.204(44)\cr
0.1670~|Wilson~|-1.28(48)~|8.1(18)~|-0.158(33)
\endtable
}
\fi

The charge radius of the dynamic staggered proton at $\kappa=0.1565$
is $\langle(r/a)^2_p\rangle=10.5\pm1.2$ while the dynamic Wilson
proton has a charge radius of $\langle(r/a)^2_p\rangle=8.1\pm1.8$. The
30\% difference in lattice spacings derived from the rho mass
translates into a ratio of $1.69$ for the charge radii.  The ratio of
the calculated charge radii is $1.30\pm0.34$, slightly below the
anticipated value.

The baryon wave functions calculated on dynamic staggered fermion
lattices are not substantially different from the quenched baryon wave
functions. The size of the proton calculated with minimal Fock space
wave functions is half the size of the physical proton, using quenched
or dynamic staggered lattices.  The baryon wave functions calculated
on dynamic Wilson fermion lattices are smaller (in lattice units) than
their counterparts in the other two formulations. The magnitude of the
difference in size is consistent with the difference in lattice
spacings which results from fixing the lattice spacings to the rho
mass.

\head{V. Conclusions}

No dramatic differences are seen between wave functions in the
quenched approximation and wave functions in full QCD. The second
moments of the pion and rho meson in lattice units are $3\sigma$ and
${3\over2}\sigma$ smaller than on the quenched lattices, respectively.
We believe most of this difference can be accounted for by a rescaling
of the lattice spacing of about 8\%.

The wave functions calculated on dynamic Wilson fermion lattices are
substantially smaller in lattice units than the corresponding wave
functions in the other two formulations. When we convert to physical
units the size of any of the particles is roughly independent of the
formulation of the lattice.

The ratio of the charge radius of the neutron to that of the proton
with staggered dynamic fermions is consistent with the experimental
ratio, as was found in the quenched simulation. In contrast, the sizes
of the particles derived from the wave functions in all three
simulations are smaller than the physical states, smaller by around a
factor of two. A more desirable method of calculating radial moments
and charge radii may be to use structure functions rather than wave
functions, as discussed in Ref.~\cite{SWAVE}.

The rho meson wave function in Coulomb gauge is larger than that for
the pion, proton or delta. This would lead one to expect that finite
size effects in spectroscopy studies would be greatest for the rho.
That this is not true is another indication that some physics which
impacts spectroscopy is not included in the minimum Fock space wave
function.

\head{Acknowledgments}
This work was supported by the U.~S. Department of Energy under contracts
DE--FG02--85ER--40213, %   Doug
DE--AC02--86ER--40253, %  Tom
DE--AC02--84ER--40125, %Steve
W-31-109-ENG-38, % Argonne
and by the National Science Foundation under grants
NSF-PHY87-01775, %  jbkogut
NSF-PHY89-04035  %  bob sugar, ITP
and
NSF-PHY86-14185. %  bob sugar
The computations were carried out at the Florida State University
Supercomputer Computations Research Institute which is partially
funded by the U.S. Department of Energy through Contract No.
DE-FC05-85ER250000 and at the Pittsburgh Supercomputing Center.  We
thank T. Kitchens and J. Mandula for their continuing support and
encouragement.

%flush any remaining figures before starting next section
\vfill\supereject

\if\preprint N
\figurecaptions
\item{1.}
{Comparison of quenched Wilson spectroscopy at $6/{g^2}=6.0$ (squares)
with Wilson valence spectroscopy from a dynamical staggered fermion
simulation at at $6/{g^2}=5.6$ (octagons) and dynamical Wilson fermion
simulation at $6/{g^2}=5.3$ (diamond).}

\item{2.}
{Coulomb gauge wave functions at time $t=6$, for separations of
$(x,0,0)$.  For the baryons the two like-flavor quarks are pinned to
the same site while the non-like quark is at separation $(x,0,0)$
relative to the other two. The meson data has been normalized after
averaging. The baryon data has been normalized on a lattice-by-lattice
basis.  Particles are labeled by boxes for pion, diamonds for rho,
octagons for proton and crosses for delta.  (a)---(c) are simulations
with dynamical staggered quarks and Wilson valence quarks, at Wilson
hopping parameters of $\kappa=0.1565, 0.1585$ and $0.1600$,
respectively. (d) is with Wilson dynamical and valence quarks at
$\kappa=0.1670$.  }

\item{3.}
{Pion wave function on lattices with staggered fermions at
all three $\kappa$ values ($\kappa=0.1565,0.1585,0.1600$) and with
Wilson fermions (at $\kappa=0.1670$).}

\item{4.}
{Correlated fit parameters for meson wave
functions (as in Eqn.~\mesonwf) from correlated fits to the data. (a)
exponential falloff $x_2$, (b) exponent $x_3$.}

\item{5.}
{Second moment of mesons (as in Eqn.~\momentint), using
parameterized wave functions of Eqn.~\mesonwf.  Crosses indicate
staggered dynamic fermions, squares indicate Wilson dynamic fermions,
diamonds are quenched.  (a) pion, (b) rho.  }

\item{6.}
{Meson second moments as calculated through discrete lattice
sums, via Eqn.~\discretemoment.  Crosses indicate staggered dynamic
fermions, squares indicate Wilson dynamic fermions, diamonds are
quenched. (a) pion, (b) rho.}

\item{7.}
{Falloff of wave function of $b$ quark (crosses) with
separation from an $aa$ diquark, and of an $a$ quark from an $ab$
diquark (octagons). (a) nucleon, (b) delta. Data are at
$\kappa=0.1585$, with dynamic staggered fermions.}

\item{8.}
{Nucleon wave function data points at $\kappa=0.1585$ (with dynamic
staggered fermions) with functional forms of Eqn.~\psiaab\ and
Eqn.~\psiaba\ overplotted.  Crosses represent wave function for quark
of flavor $b$ relative to $aa$ diquark, octagons represent wave
function for quark of flavor $a$ relative to $ab$ diquark.}

\item{9.}
{Exponential falloffs $x_{aa}$ and $x_{ab}$ which
parameterize the nucleon wave function, as in Eqn.~\proposepsi. }

\item{10.}
{Exponential falloffs $x_{aa}$ and $x_{ab}$ which
parameterize the wave function of the delta, as in Eqn.~\proposepsi. }

\item{11.}
{Value of the exponent $y$ which parametrizes the baryon
wave functions, as in Eqn.~\proposepsi. (a) nucleon, (b) delta. }

\item{12.}
{Ratio of charge radii of neutron to proton as a function of
pion mass. Horizontal dashed line is the experimental ratio. Crosses
indicate staggered dynamic fermions, squares indicate Wilson dynamic
fermions, diamonds are quenched. }

\endfigurecaptions

\head{Table Captions}

\item{1.} Correlated Fit Parameters for Mesons

\item{2.} Correlated Fit Function Parameters for Baryons

\item{3.} Charge Radii of Baryons

\head{Tables}

\vbox{
\bigskip
\centerline{ TABLE~\dmesxt}
\bigskip
\begintable
$\kappa$~|sea quarks~|pion $x_2$~|pion $x_3$~|rho $x_2$~|rho $x_3$\cr
0.145~|quenched~|0.2069(16)~|1.274(9)~|0.0972(14)~|1.514(11)~\cr
0.152~|quenched~|0.1970(10)~|1.247(7)~|0.0727(12)~|1.534(15)~\cr
0.153~|quenched~|0.1964(9)~|1.241(7)~|0.0698(14)~|1.534(17)~\cr
0.154~|quenched~|0.1961(9)~|1.234(7)~|0.0674(18)~|1.532(21)~\cr
0.155~|quenched~|0.1960(10)~|1.228(8)~|0.0658(26)~|1.529(28)~\cr
0.1565~|staggered~|0.234(2)~|1.216(7)~|0.094(3)~|1.45(2)~\cr
0.1585~|staggered~|0.232(2)~|1.211(8)~|0.086(4)~|1.46(3)~\cr
0.1600~|staggered~|0.234(2)~|1.213(11)~|0.080(5)~|1.48(3)~\cr
0.1670~|Wilson~|0.337(5)~|1.253(9)~|0.121(4)~|1.58(2)
\endtable
}

\vbox{
\bigskip
\centerline{ TABLE~\dxy }
\centerline{(a): Nucleon}
\bigskip
\begintable
$\kappa$~|sea quarks~|$x_a$~|$x_b$~|$x_{aa}$~|$x_{ab}$~|y\cr
0.145~|quenched~|0.158(3)~|0.177(3)~|0.069(3)~|0.089(1)~|1.398(12)\cr
0.152~|quenched~|0.135(3)~|0.156(2)~|0.057(3)~|0.078(1)~|1.378(11)\cr
0.153~|quenched~|0.132(3)~|0.153(2)~|0.055(3)~|0.077(1)~|1.372(12)\cr
0.154~|quenched~|0.129(3)~|0.151(3)~|0.054(4)~|0.076(1)~|1.365(14)\cr
0.155~|quenched~|0.125(4)~|0.149(3)~|0.050(5)~|0.074(2)~|1.361(19)\cr
0.1565~|staggered~|0.158(4)~|0.187(4)~|0.065(4)~|0.094(2)~|1.363(12)\cr
0.1585~|staggered~|0.148(4)~|0.180(4)~|0.058(5)~|0.090(2)~|1.355(12)\cr
0.1600~|staggered~|0.123(10)~|0.170(5)~|0.038(10)~|0.085(3)~|1.35(4)\cr
0.1670~|Wilson~|0.208(9)~|0.264(8)~|0.076(10)~|0.132(4)~|1.33(3)
\endtable
\bigskip
\centerline{(b): Delta}
\bigskip
\begintable
$\kappa$~|$x_a$~|$x_b$~|$x_{aa}$~|$x_{ab}$~|y\cr
0.145~|0.150(3)~|0.141(3)~|0.079(3)~|0.071(1)~|1.410(13)\cr
0.152~|0.124(3)~|0.116(2)~|0.066(3)~|0.058(1)~|1.376(15)\cr
0.153~|0.122(3)~|0.113(2)~|0.065(4)~|0.057(1)~|1.367(17)\cr
0.154~|0.120(4)~|0.111(2)~|0.064(4)~|0.056(1)~|1.357(19)\cr
0.155~|0.119(6)~|0.109(3)~|0.064(6)~|0.055(2)~|1.346(26)\cr
0.1565~|0.159(6)~|0.155(5)~|0.081(6)~|0.078(3)~|1.323(23)\cr
0.1585~|0.152(8)~|0.154(6)~|0.075(9)~|0.077(3)~|1.31(3)\cr
0.1600~|0.159(20)~|0.142(9)~|0.088(21)~|0.071(4)~|1.26(6)\cr
0.1670~|0.217(17)~|0.226(26)~|0.104(22)~|0.113(13)~|1.28(11)
\endtable}

\vbox{
\bigskip
\centerline{ TABLE~\dnuccr}
\bigskip
\begintable
$\kappa$~|sea quarks~|$\langle (r/a)^2_n\rangle$~|
$\langle (r/a)^2_p\rangle$~|
$\langle r^2_n\rangle/\langle r^2_p\rangle$\cr
0.145~|quenched~|-0.73(38)~|9.1(12)~|-0.080(34)\cr
0.152~|quenched~|-1.27(59)~|12.4(18)~|-0.102(37)\cr
0.153~|quenched~|-1.37(64)~|13.1(20)~|-0.104(38)\cr
0.154~|quenched~|-1.55(71)~|14.0(23)~|-0.111(38)\cr
0.155~|quenched~|-1.83(82)~|15.0(29)~|-0.122(38)\cr
0.1565~|staggered~|-1.23(33)~|10.5(12)~|-0.117(23)\cr
0.1585~|staggered~|-1.63(43)~|12.1(15)~|-0.135(24)\cr
0.1600~|staggered~|-3.5(18)~|17.4(61)~|-0.204(44)\cr
0.1670~|Wilson~|-1.28(48)~|8.1(18)~|-0.158(33)
\endtable
}

\fi

\references

\refis{APE}
S. Cabasino et. al., {\it Phys. Lett.} {\bf B258}, 195 (1991).

\refis{LIU}
C. Liu, in
 the Proceedings of Lattice '90,
{\sl Nucl. Phys.} {\bf B (Proc. Suppl) 20},  (1991) 149.
 A. D. Kennedy, \journal Intl. J. Mod. Phys., C3, 1, 1992.

\refis{OVERRELAX}
J. E.~Mandula and M. C.~Ogilvie, \pl B248, 156, 1990.

\refis{LATESTGC}
K.~M. Bitar, et.~al., Colorado preprint COLO-HEP-278 (April 1992);
Phys. Rev. D, in press.

\refis{GCW}
K.~M. Bitar, et.~al., in preparation.

\refis{HMC}
S. Duane, A. Kennedy, B. Pendleton, and D. Roweth, \pl 194B, 271, 1987.

\refis{DEGRANDILU}
T. DeGrand, \journal Comput. Phys. Commun., 52, 161, 1988.

\refis{HEMCGC}
K.~Bitar et al., \prl 65, 2106, 1990, \prd 42, 3794, 1990.

\refis{RELAX}
F. Brown and T. Woch, \journal Phys. Rev. Lett., 58, 2394, 1987 ; M.
Creutz, \journal Phys. Rev., D36, 55, 1987 .  For a review, see S.
Adler {\sl Nucl. Phys.} {\bf B (Proc. Suppl) 9}, (1989) 437.

\refis{HMD}
H.~C.~Andersen, \journal J. Chem. Phys., 72, 2384, 1980;
S. Duane, \np B257, 652, 1985;
S. Duane and J. Kogut, \prl 55, 2774, 1985;
S. Gottlieb, W. Liu, D. Toussaint, R. Renken and
R. Sugar, \prd 35, 2531, 1987.

\refis{WILSONWAVES}
See, for example, K. C. Bowler, et. al.,
\journal Nucl. Phys., B240, 213, 1984.

\refis{KPQ}
A. Kennedy and B. Pendleton, \journal Phys. Lett., 156B, 393, 1985.

\refis{DOITRIGHT}
For a good introduction to error analysis see
D. Toussaint, in
``From Actions to Answers--Proceedings of the 1989 Theoretical Advanced
Summer Institute in Particle Physics,'' T. DeGrand and D. Toussaint, eds.,
(World, 1990).

\refis{GOTT}
S. Gottlieb, in ``Advances in Lattice Gauge Theory,''
D. Duke and J. Owens, eds. (World Scientific, 1985).

\refis{VELIK}
B.
Velikson and D. Weingarten, \journal Nucl. Phys., B249, 433, 1985

\refis{NEGELE}
 M.-C.~Chu, M.~Lassia and J. W.~Negele, \journal Nucl. Phys., B360, 31,  1991.

\refis{FNAL}
A. El-Khadra, G. Hockney, A. Kronfeld, and P. Mackenzie,
Fermilab preprint PUB-91/354-T (1991).

\refis{HL}
C. Bernard, J. Labrenz, and A. Soni, Nucl. Phys.
{\bf B (Proc. Suppl.)  20}, 488 (1991).

\refis{PWAVE}
T. DeGrand and M. Hecht,
\journal Phys. Lett., B275, 435, 1992.

\refis{SWAVE}
M. Hecht and T. DeGrand, Colorado preprint COLO-HEP-277 (1992);
{\it Phys. Rev.} {\bf D}, in press.

\refis{MILC}
C. Bernard, et.~al.,
\journal Phys. Rev. Lett., 68, 2125, 1992.

\refis{DALLY}
E.~B.~Dally et. al., \prl 48, 375, 1982.

\refis{CHARGERATIO}
V.~E. Krohn and G.~R. Ringo,
\prd 8, 1305, 1973;
 R.~W. Berard et. al.,
\pl B47, 355, 1973;
F. Borkowski et.~al.,
\journal Nucl. Phys., A222, 269, 1974.

\refis{CRPROTON}
F. Borkowski et.~al.,
\journal Nucl. Phys., A222, 269, 1974;
 L.~N. Hand, D.~G. Miller and R. Wilson, \journal Revs. Mod. Phys., 35, 335,
1963.

\refis{PMASS}
{Using data published in Ref.~\cite{APE}\ and performing a linear
least squares fit of the proton mass against $(\kappa_c-\kappa)$.}

%\refis{BR}
%{G. E. Brown and M. Rho, {\it Phys. Lett.} {\bf 82B}, 177 (1979);
%G. E. Brown, M. Rho, and V. Vento, {\it Phys. Lett.} {\bf 84B}, 383
%(1979).}

%\refis{VRNJB}
%{V. Vento, M. Rho, E. Nyman, J. Jun and G. E. Brown,
%{\it Nucl. Phys.} {\bf A345}, 43 (1980).}

%\refis{BRW}
%{G. E. Brown, M. Rho and W. Weise, {\it Nucl. Phys.} {\bf A454}, 669
%(1986).}

%\refis{BANDM}
%{V. Bernard and U-G. Meissner, {\it Phys. Rev. Lett.} {\bf 61}, 2296
%(1988).}

%\refis{FMY}
%{F. Myhrer, {\it Phys. Lett.} {\bf 110B}, 353 (1982).}

%\refis{PANDD}
%{T.~A. DeGrand and M.~W. Hecht, work in progress.}

\endreferences
\endit
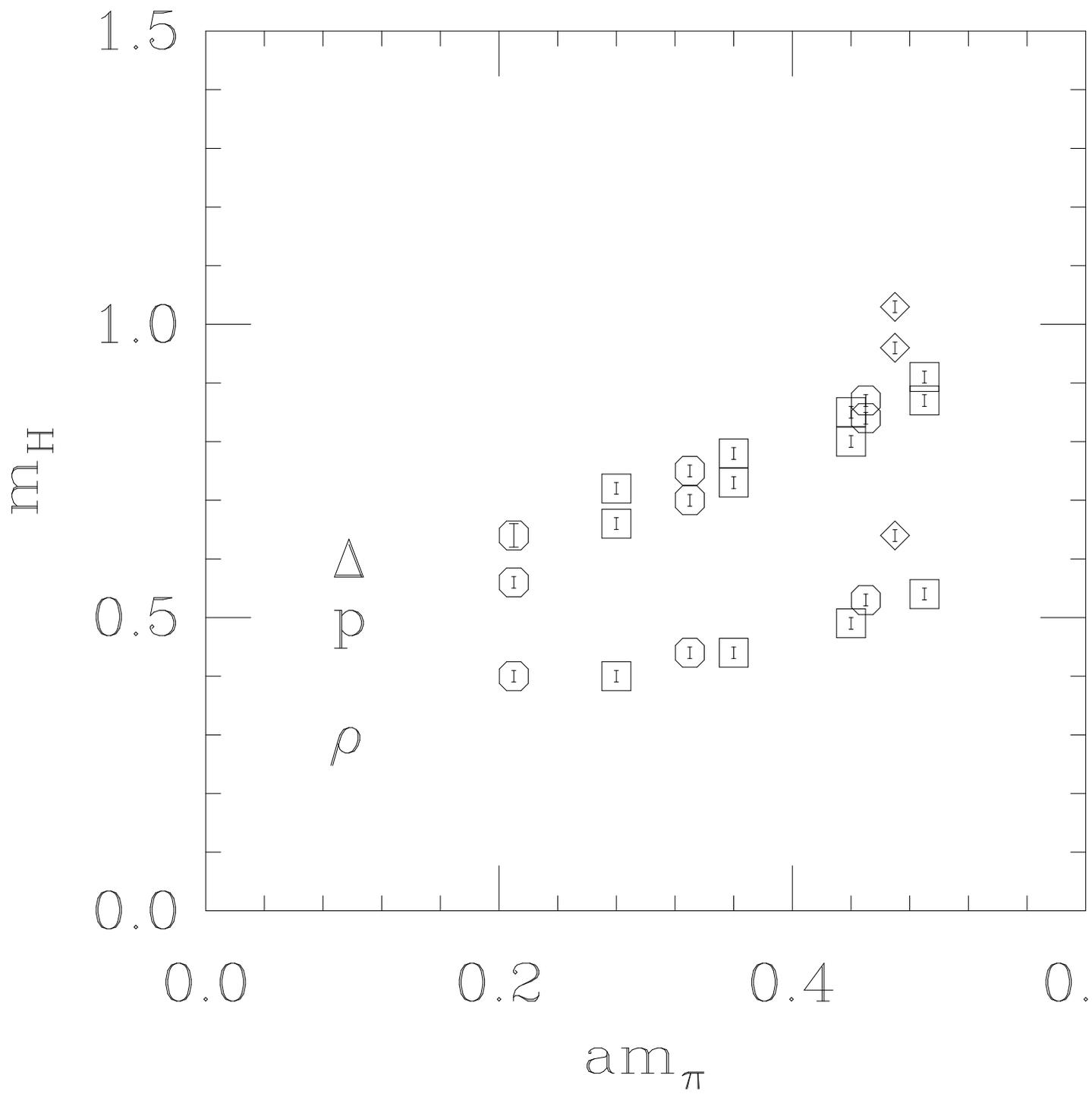
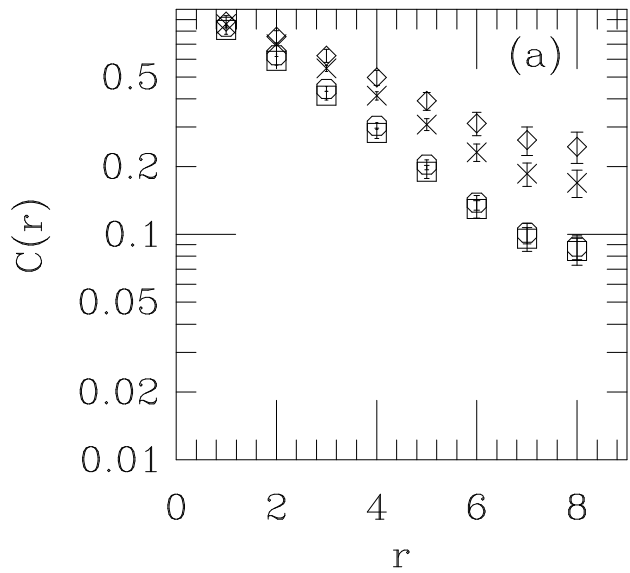
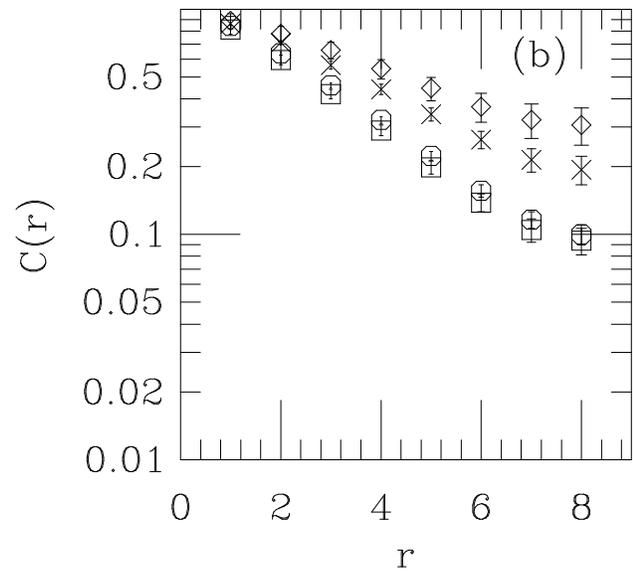
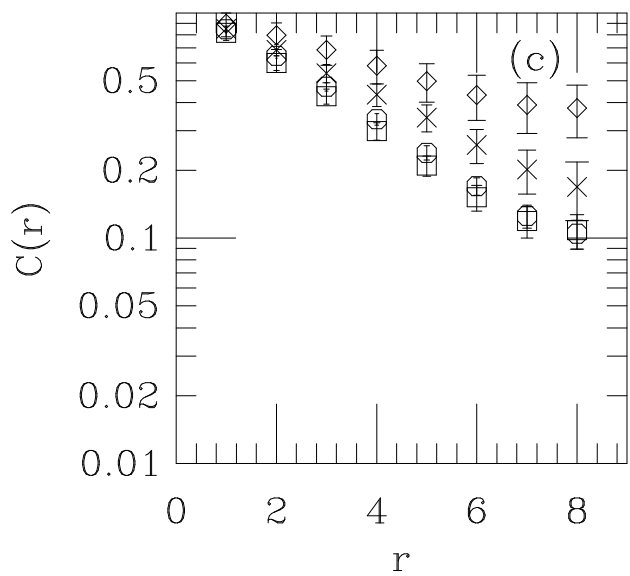
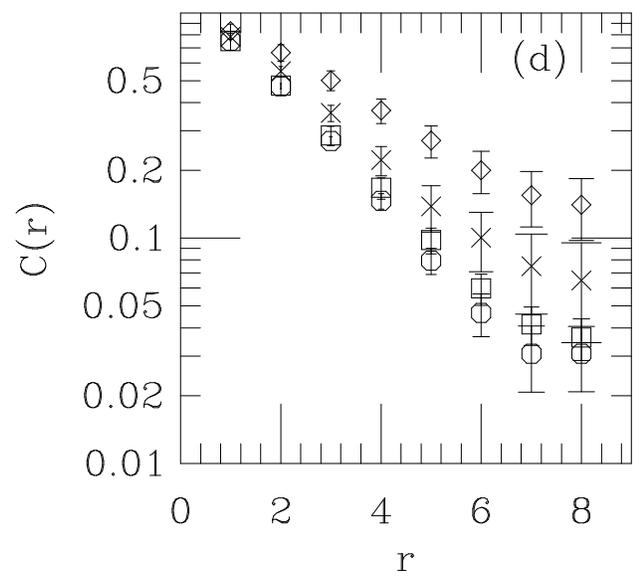
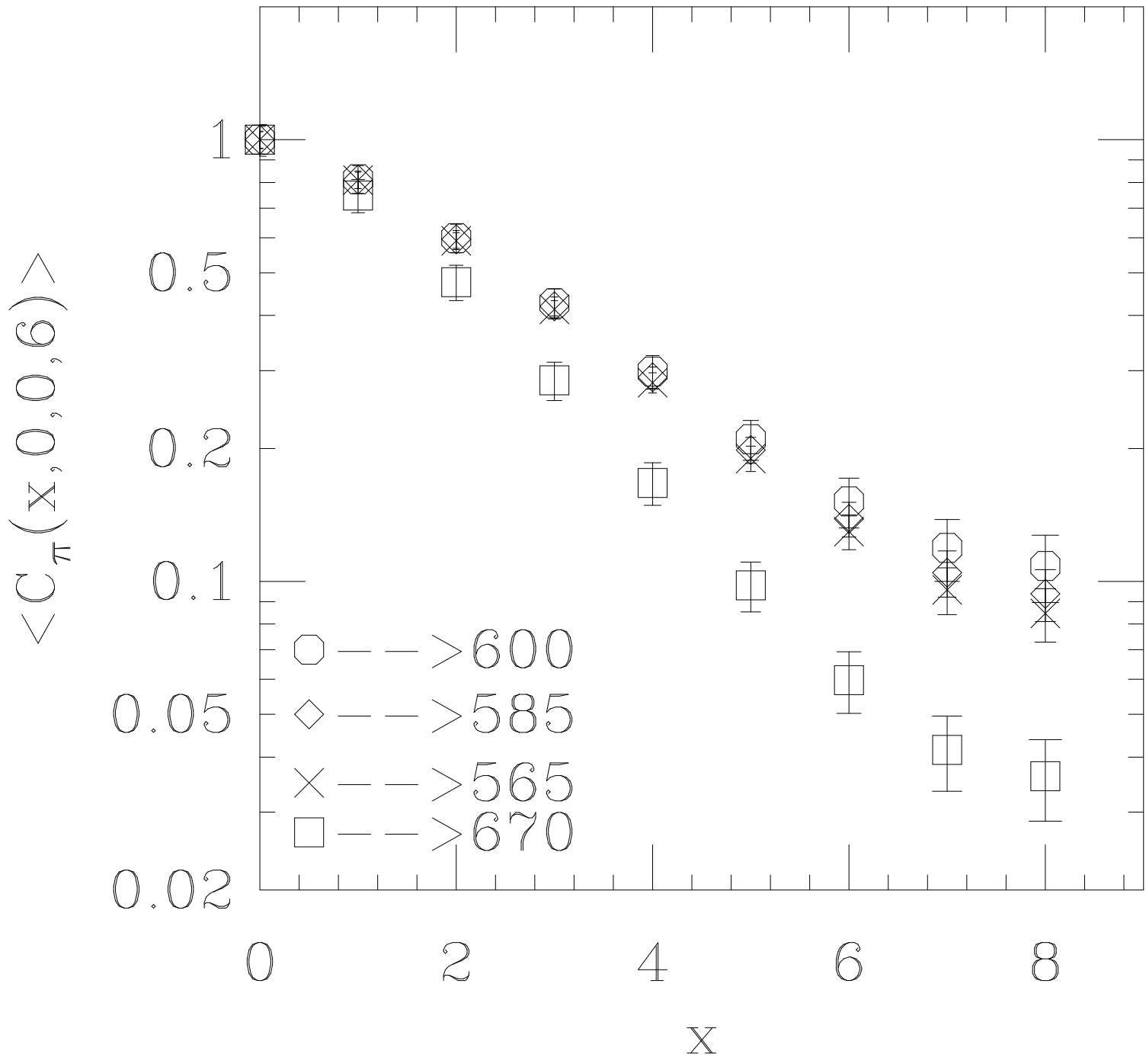
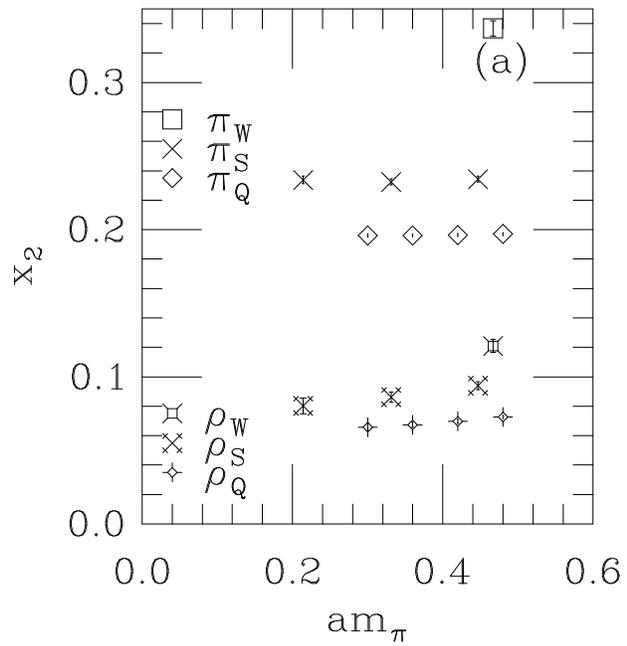
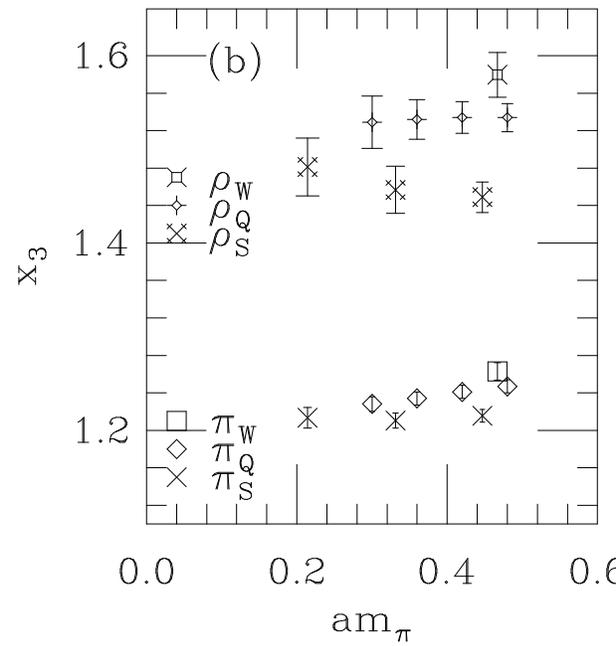
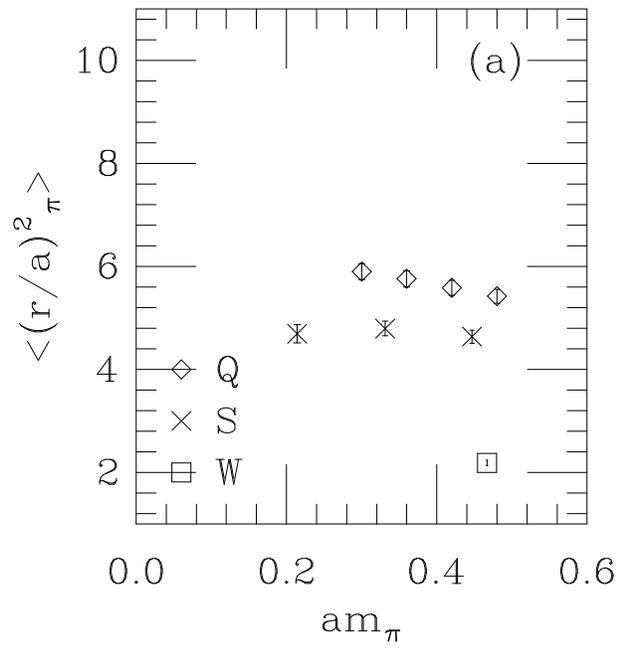
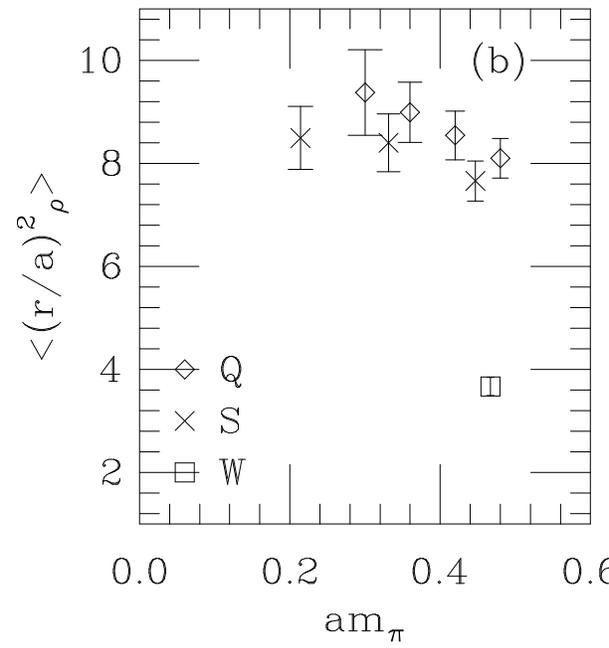
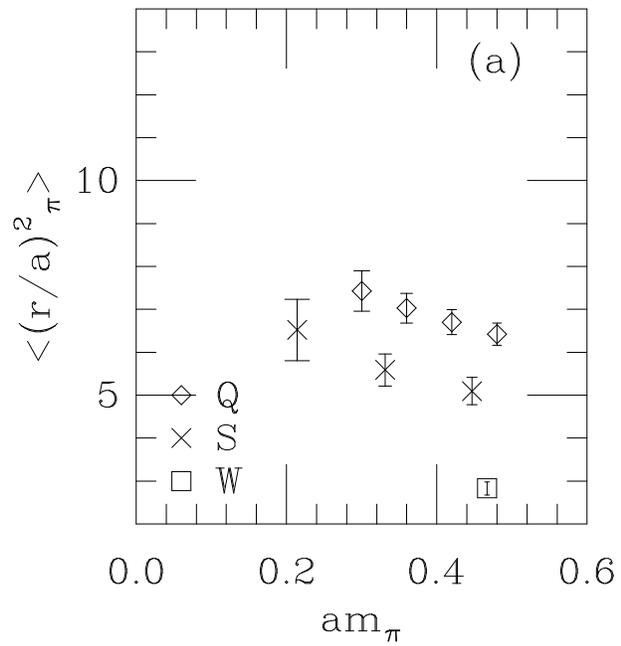
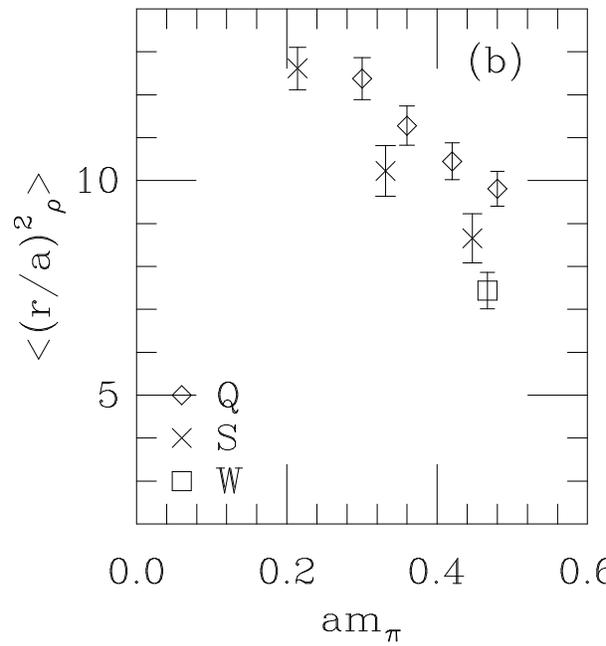
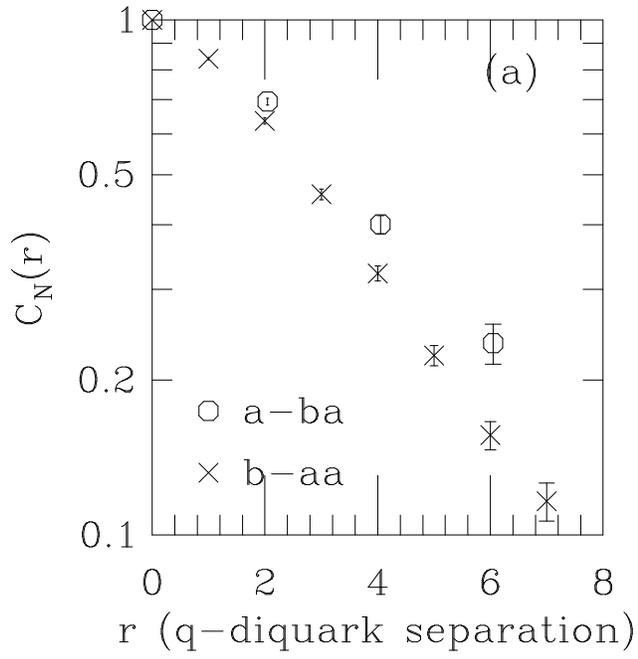
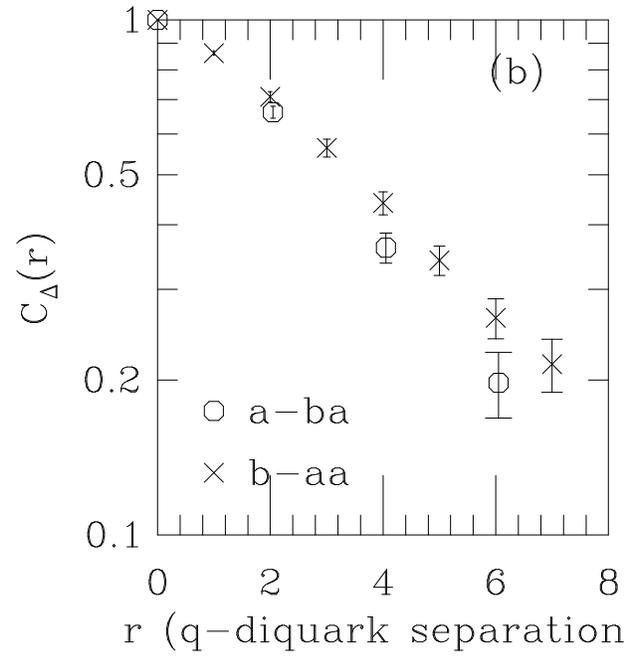
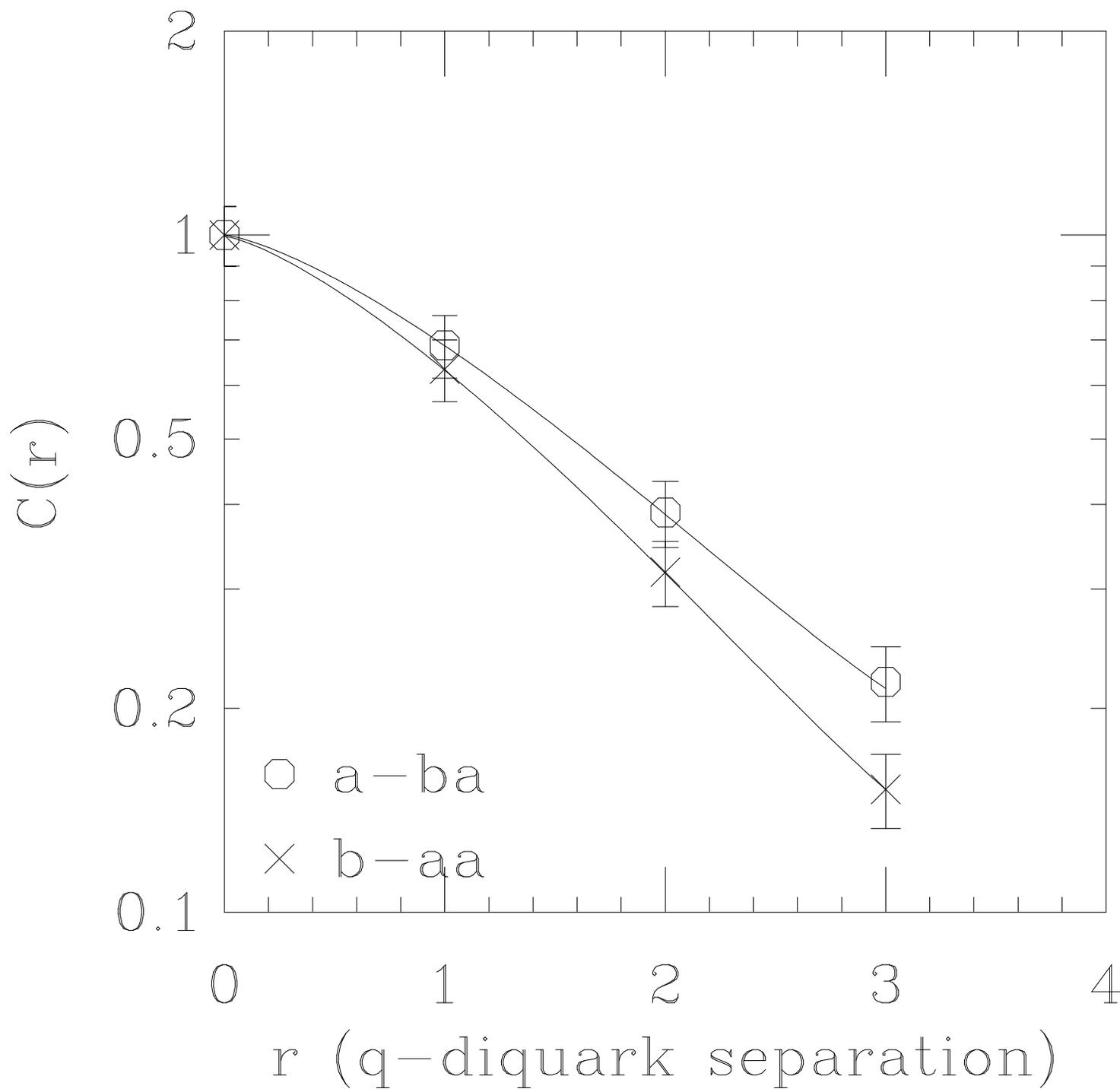
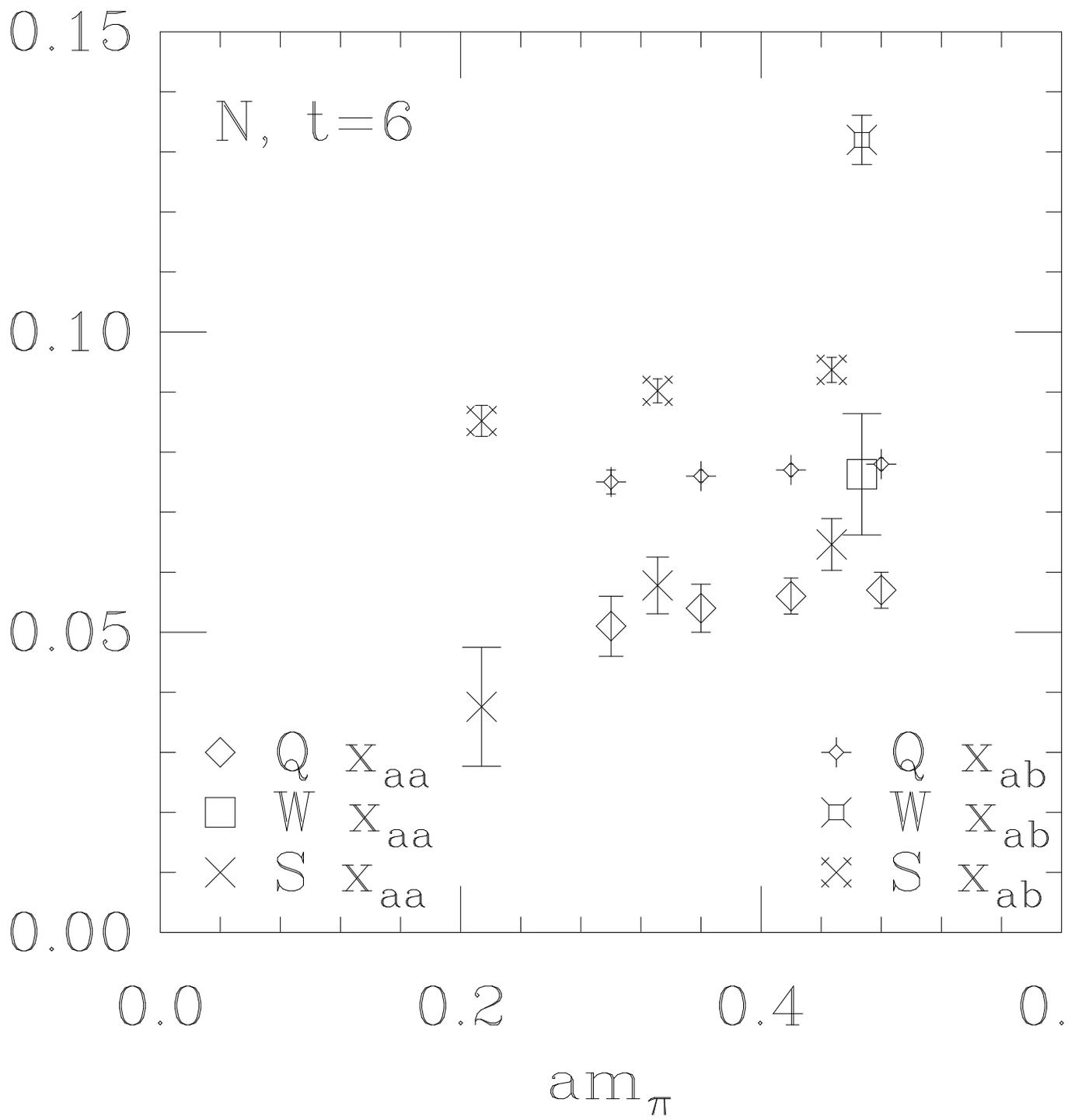
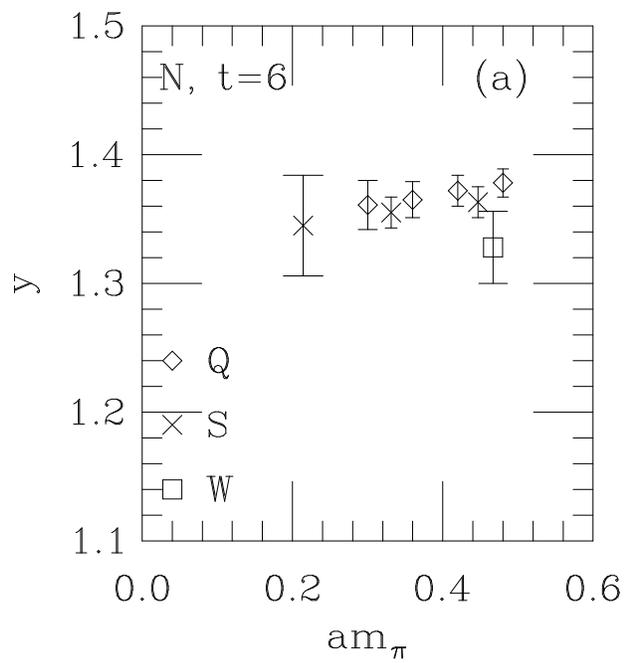
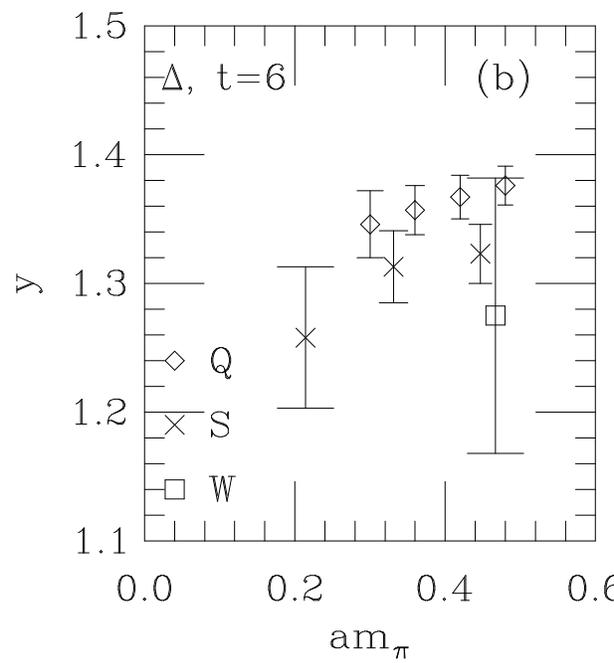
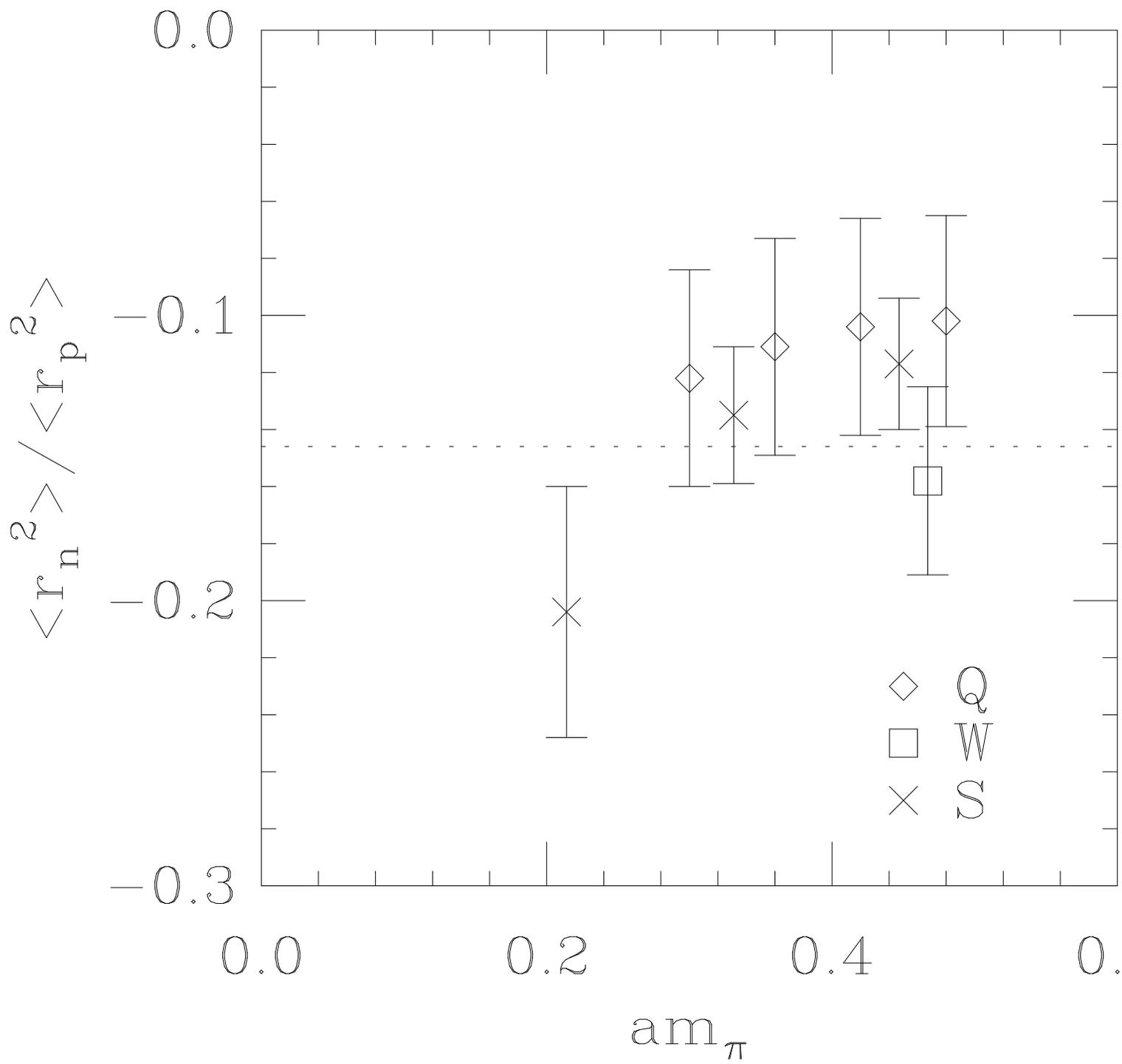
